\documentstyle[preprint,tighten,eqsecnum,aps,floats,psfig]{revtex}

\def\ub{{\overline{u}}}
\def\vb{{\overline{v}}}

\begin{document}
\draft

\title{
Randomly dilute spin models: a six-loop field-theoretic study.
}
\author{Andrea Pelissetto$\,^1$ and 
Ettore Vicari$\,^2$ }
\address{$^1$ Dipartimento di Fisica dell'Universit\`a di Roma I
and I.N.F.N., I-00185 Roma, Italy.}
\address{$^2$
Dipartimento di Fisica dell'Universit\`a 
and I.N.F.N., 
Via Buonarroti 2, I-56127 Pisa, Italy.\\
{\bf e-mail: \rm
{\tt Andrea.Pelissetto@roma1.infn.it},
{\tt vicari@mailbox.difi.unipi.it}
}}

\date{\today}

\maketitle

\begin{abstract}
We consider the Ginzburg-Landau $MN$-model that describes 
$M$ $N$-vector cubic models with $O(M)$-symmetric couplings.
We compute the 
renormalization-group functions to six-loop order in $d=3$.
We focus on the limit $N\to 0$ which describes the critical 
behaviour of an $M$-vector model in the presence of weak quenched
disorder. We perform a detailed analysis of the perturbative 
series for the random Ising model $(M=1)$. We obtain for the 
critical exponents: $\gamma = 1.330(17)$, $\nu = 0.678(10)$,
$\eta = 0.030(3)$, $\alpha=-0.034(30)$, $\beta = 0.349(5)$,
$\omega = 0.25(10)$. For $M\ge 2$ we show that the $O(M)$ fixed
point is stable, in agreement with general non-perturbative 
arguments, and that no random fixed point exists.
\end{abstract}



\section{Introduction.}

The critical behavior of systems with quenched disorder is of considerable 
theoretical and experimental interest. 
A typical example is obtained by mixing an (anti)-ferromagnetic 
material with a non-magnetic one, obtaining the so-called dilute 
magnets. These materials are usually described in terms of 
a lattice short-range Hamiltonian of the form
\begin{equation}
{\cal H}_x = - J\,\sum_{<ij>}  \rho_i \,\rho_j \;\vec{s}_i \cdot \vec{s}_j,
\label{latticeH}
\end{equation}
where $\vec{s}_i$ is an $M$-component spin and 
the sum is extended over all nearest-neighbor sites. The quantities
$\rho_i$ are uncorrelated random variables, which are equal to one 
with probability $x$ (the spin concentration) and zero with probability $1-x$
(the impurity concentration). The pure system corresponds to $x=1$.
One considers quenched disorder, since the relaxation time associated to 
the diffusion of the impurities is much larger than all other typical time 
scales, so that, for all practical purposes, one can consider the position
of the impurities fixed.
For sufficiently low spin dilution $1-x$, i.e. as long as one is above the 
percolation threshold of the magnetic atoms.
the system described by the Hamiltonian ${\cal H}_x$ undergoes a second-order 
phase transition at $T_c(x) < T_c(x=1)$
(see e.g. Ref.~\cite{Stinchcombe-83} for a review).

The relevant question in the study of this class of systems is the effect
of the disorder on the critical behavior. 
The Harris criterion~\cite{Harris-74} states that the addition of 
impurities to a system which undergoes a second-order 
phase transition does not change the critical behavior 
if the specific-heat critical exponent $\alpha_{\rm pure}$ of the pure 
system is negative. If $\alpha_{\rm pure}$ is positive, the transition
is altered. Indeed, in disordered systems the exponent $\nu$ satisfies the 
inequality $\nu\geq 2/d$ \cite{CCFS-86,AHW-98} 
--- this fact has been questioned however
in Refs.~\cite{SF-88,PSZ-97} --- and therefore, by hyperscaling,
$\alpha_{\rm random}$ is negative. Thus, if $\alpha_{\rm pure}$ is positive, 
$\alpha_{\rm random}$ differs from $\alpha_{\rm pure}$, so that
the pure system and the dilute one have a different critical behavior.
In pure $M$-vector models with $M>1$, the specific-heat exponent 
$\alpha_{\rm pure}$ is negative; therefore, according to the Harris criterion, 
no change in the critical asymptotic behavior is expected
in the presence of weak quenched disorder. This means that in these systems
disorder leads only to irrelevant scaling corrections.
Three-dimensional Ising systems are more interesting, since 
$\alpha_{\rm pure}$ is positive.
In this case, the presence of quenched impurities leads
to a new universality class.

Theoretical investigations, using approaches based on the renormalization 
group 
\cite{HL-74,Emery-75,Lubensky-75,Khmelnitskii-75,AIM-76,GL-76,%
Aharony-76,GMM-77,JK-77,Shalaev-77,NR-82,Jug-83,Newlove-83,%
MS-84,PA-85,DG-85,Shpot-89,MSS-89,Mayer-89,Shpot-90,HS-92,BS-92,JOS-95,%
SAS-97,FHY-98,HY-98,Mayer-98,FHY-99,Varnashev-99,FHY-00,PS-99},
and 
numerical Monte Carlo simulations 
\cite{Landau-80,MLT-86,CS-86,BS-88,Betal-89,WS-89,WWMC-90,%
HF-90,Heuer-90,WHF-92,Heuer-93,Hennecke-93,WD-98,BFMMPR-98,Hukushima-2000},
support the existence of a new random Ising fixed point describing the 
critical behavior along the $T_c(x)$ line: the critical exponents are 
dilution independent
(for sufficiently low dilution) and different from those of the 
pure Ising model.

Experiments confirm this picture. 
Cristalline mixtures of an Ising-like
uniaxial antiferromagnet with short-range
interactions (e.g. FeF${}_2$, MnF${}_2$) with a nonmagnetic material
(e.g. ZnF${}_2$ )
provide a typical realization of the random Ising model (RIM)
(see e.g. Refs. 
\cite{DG-81,BCSJBKJ-83,HCK-85,BKJ-86,Barret-86,MCYBUB-86,%
BKFJ-88,RKLHE-88,RKJ-88,TPBH-88,FKJ-91,%
BWSHNLRL-95,BWSHNLRL-96,HFHBRC-97,SBF-98,SB-98,SBF-99}).
Some experimental results are reported in Table~\ref{experiments}. 
This is not a complete list, but it gives 
an overview of the experimental state of the art. 
Other experimental results can be found in Refs.~\cite{Stinchcombe-83,FHY-00}.
The experimental estimates are definitely different from the values of the 
critical exponents for pure Ising systems, which are 
(see Ref. \cite{CPRV-99} and references therein)
$\gamma=1.2371(4)$, $\nu=0.63002(23)$, $\alpha=0.1099(7)$, and 
$\beta=0.32648(18)$.
Moreover, they appear to be independent of the concentration.
We mention that in the presence of an external magnetic field along the 
uniaxial direction,
dilute Ising systems present a different critical behavior, 
equivalent to that of the random-field Ising model~\cite{FA-79}.
This is also the object of intensive theoretical and experimental 
investigations (see e.g. Refs.~\cite{Belanger-98,Nattermann-98}).

Several experiments also tested the effect of disorder 
on the $\lambda$-transition of ${}^4$He that belongs to the 
$XY$ universality class, corresponding to $M=2$
\cite{KHR-75,FGWC-88,CBMWR-88,TCR-92,YC-97,ZR-99}. They studied
the critical behaviour of ${}^4$He completely filling the pores 
of porous gold or Vycor glass. The results indicate that the 
transition is in the same universality class of the 
$\lambda$-transition of the pure system in agreement with the 
Harris criterion \cite{footnoteHedilute}.

\begin{table}[tbp]
\caption{
Experimental estimates of the critical exponents for systems in the 
RIM universality class.
}
\label{experiments}
\begin{tabular}{ccccccc}
\multicolumn{1}{c}{Ref.}&
\multicolumn{1}{c}{material}&
\multicolumn{1}{c}{concentration}&
\multicolumn{1}{c}{$\gamma$}&
\multicolumn{1}{c}{$\nu$}&
\multicolumn{1}{c}{$\alpha$}&
\multicolumn{1}{c}{$\beta$}\\
\tableline \hline
\cite{BKJ-86} & Fe${}_x$Zn${}_{1-x}$F${}_2$ & 
                $x=0.46$ & $1.33(2)$ & $0.69(3)$ & & \\
\cite{MCYBUB-86} & Mn${}_x$Zn${}_{1-x}$F${}_2$ & 
                $x=0.75$ & $1.364(76)$ & $0.715(35)$ & & \\
\cite{RKLHE-88}  & Fe${}_x$Zn${}_{1-x}$F${}_2$ & 
               $x=0.9$  &                  &                  & & $0.350(9)$ \\
\cite{RKJ-88}    & Mn${}_x$Zn${}_{1-x}$F${}_2$ & 
                $x=0.40,0.55,0.83$  &                  &       & $-0.09(3)$& \\
\cite{TPBH-88}   & Mn${}_x$Zn${}_{1-x}$F${}_2$ & 
                $x=0.5$  &                  &                  & & $0.33(2)$ \\
\cite{HFHBRC-97} & Fe${}_x$Zn${}_{1-x}$F${}_2$ & 
                $x=0.5$  &                  &                  & & $0.36(2)$ \\
\cite{SB-98}     & Fe${}_x$Zn${}_{1-x}$F${}_2$ & 
                $x=0.93$ &                  &       & $-0.10(2)$& \\
\cite{SBF-99} &  Fe${}_x$Zn${}_{1-x}$F${}_2$ & 
                $x=0.93$ & $1.34(6)$ & $0.70(2)$ & & \\
\end{tabular}
\end{table}

The starting point of the field-theoretic approach to the study of 
ferromagnets in the presence of quenched disorder
is the Ginzburg-Landau-Wilson Hamiltonian~\cite{GL-76}
\begin{equation}
{\cal H} = \int d^d x 
\left\{ {1\over 2}(\partial_\mu \phi(x))^2 + {1\over 2} r \phi(x)^2 + 
{1\over 2} \psi(x) \phi(x)^2 + 
{1\over 4!} g_0 \left[ \phi(x)^2\right]^2 \right\},
\label{Hphi4ran}
\end{equation}
where $r\propto T-T_c$, and $\psi(x)$ is a spatially uncorrelated 
random field with Gaussian distribution
\begin{equation}
P(\psi) = {1\over \sqrt{4\pi} w} \exp\left[ - {\psi^2\over 4 w}\right].
\end{equation}
We consider quenched disorder. Therefore, in order
to obtain the free energy of the system, we must 
compute the partition function $Z(\psi,g_0)$
for a given distribution $\psi(x)$, and then average the corresponding
free energy over all distributions with probability 
$P(\psi)$. By using the standard replica trick, it is possible 
to replace the quenched average with an annealed one.
First, the system is replaced by $N$ non-interacting
copies with annealed disorder.
Then, integrating over the disorder, one obtains the
Hamiltonian~\cite{GL-76}
\begin{equation}
{\cal H}_{MN} = \int d^d x 
\left\{ \sum_{i,a}{1\over 2} \left[ (\partial_\mu \phi_{a,i})^2 + 
         r \phi_{a,i}^2 \right] + 
  \sum_{ij,ab} {1\over 4!}\left( u_0 + v_0 \delta_{ij} \right)
          \phi^2_{a,i} \phi^2_{b,j} 
\right\}
\label{Hphi4}
\end{equation}
where $a,b=1,...M$ and $i,j=1,...N$.
The original system, i.e.   
the dilute $M$-vector model, is recovered 
in the limit $N\rightarrow 0$. Note that 
the coupling $u_0$ is negative (being proportional to minus the variance of
the quenched disorder), while the coupling $v_0$ is positive. 

In this formulation, 
the critical properties of the dilute $M$-vector model
can be investigated by studying the renormalization-group flow
of the Hamiltonian (\ref{Hphi4}) in the limit $N\rightarrow 0$, 
i.e. of ${\cal H}_{M0}$. One can then apply
conventional computational schemes, such
as the $\epsilon$-expansion, the fixed-dimension $d=3$ expansion, 
the scaling-field method, etc...
In the renormalization-group approach, 
if the fixed point corresponding to the pure 
model is unstable and the renormalization-group flow moves
towards a new random fixed point, then the random system has a different
critical behavior.

It is important to note that in the renormalization-group approach one assumes
that the replica symmetry is not broken.
In recent years, however, this picture has been
questioned~\cite{DHSS-95,DF-95,Dotsenko-99} on the ground that 
the renormalization-group approach does note take into account other  
local minimum configurations of the random Hamiltonian (\ref{Hphi4ran}), which
may cause the spontaneous breaking of the replica symmetry.
In this paper we  assume the validity of the standard
renormalization-group approach, and simply consider the 
Hamiltonian (\ref{Hphi4}) for $N=0$.

For generic values of $M$ and $N$, the Hamiltonian ${\cal H}_{MN}$ describes
$M$ coupled $N$-vector models 
and it is usually called $MN$ model~\cite{Aharony-76}.
${\cal H}_{MN}$ has four fixed points: the trivial Gaussian one,
the O($M$)-symmetric fixed point describing $N$ decoupled $M$-vector models,
the O($MN$)-symmetric and the mixed fixed point. 
The Gaussian one is never stable. 
The stability of the other fixed points depends on the values
of $M$ and $N$ (see e.g. Ref.~\cite{Aharony-76} for a discussion).
The stability properties of the decoupled O($M$) fixed point can be 
inferred by observing
that the crossover exponent associated with the O($MN$)-symmetric interaction
(with coupling $u_0$)
is related to the specific-heat critical exponent of the O($M$) 
fixed point~\cite{Sak-74,Aharony-76}.
Indeed, at the O($M$)-symmetric fixed point one may interpret ${\cal H}_{MN}$
as the Hamiltonian of $N$ $M$-vector systems coupled by the 
O($MN$)-symmetric term. But this interaction is the sum of the products 
of the energy operators of the different $M$-vector models.
Therefore, at the O($M$) fixed point, the crossover exponent $\phi$ 
associated to the O($MN$)-symmetric quartic term should be given by
the specific-heat critical exponent $\alpha_M$ of the $M$-vector model, 
independently of $N$.
This argument implies that for $M=1$ (Ising-like systems) the pure 
Ising fixed point is unstable since $\phi = \alpha_I > 0$, 
while for $M>1$ the O($M$) fixed point is stable given that $\alpha_M<0$.
This is a general result that should hold independently of $N$. 

For the quenched disordered systems described by the 
Hamiltonian ${\cal H}_{M0}$,
the physically relevant region for the renormalization-group flow 
corresponds to negative values of 
the coupling $u$~\cite{GL-76,AIM-76}. Therefore, for 
$M>1$ the renormalization group flow is driven towards the pure O($M$) 
fixed point, and the quenched
disorder yields correction to scaling proportional to
the spin dilution and to $|t|^{\Delta_r}$ with 
$\Delta_r = - \alpha_M$.
Note that for the physically interesting two- and three-vector models
the absolute value of $\alpha_M$ is very small: 
$\alpha_2\simeq -0.013$ 
(see e.g. the recent results of Refs.~\cite{LSNCI-96,GZ-98,HT-99,CPRV-00})
and $\alpha_3\simeq - 0.12$ (see e.g. \cite{GZ-98,CPV-00}).
Thus disorder gives rise to very slowly-decaying scaling corrections.
For Ising-like systems, the pure Ising fixed point is instead unstable, 
and the flow for negative
values of the quartic coupling $u$ leads to the stable mixed or 
random fixed point which is located in the region of negative values of $u$.
The above picture emerges clearly in the framework of the $\epsilon$-expansion,
although for the Ising-like systems the RIM fixed point is 
of order $\sqrt{\epsilon}$~\cite{Khmelnitskii-75} rather than $\epsilon$.

\begin{figure}[tb]
\hspace{-2cm}
\vspace{-0.5cm}
\centerline{\psfig{width=18truecm,angle=-90,file=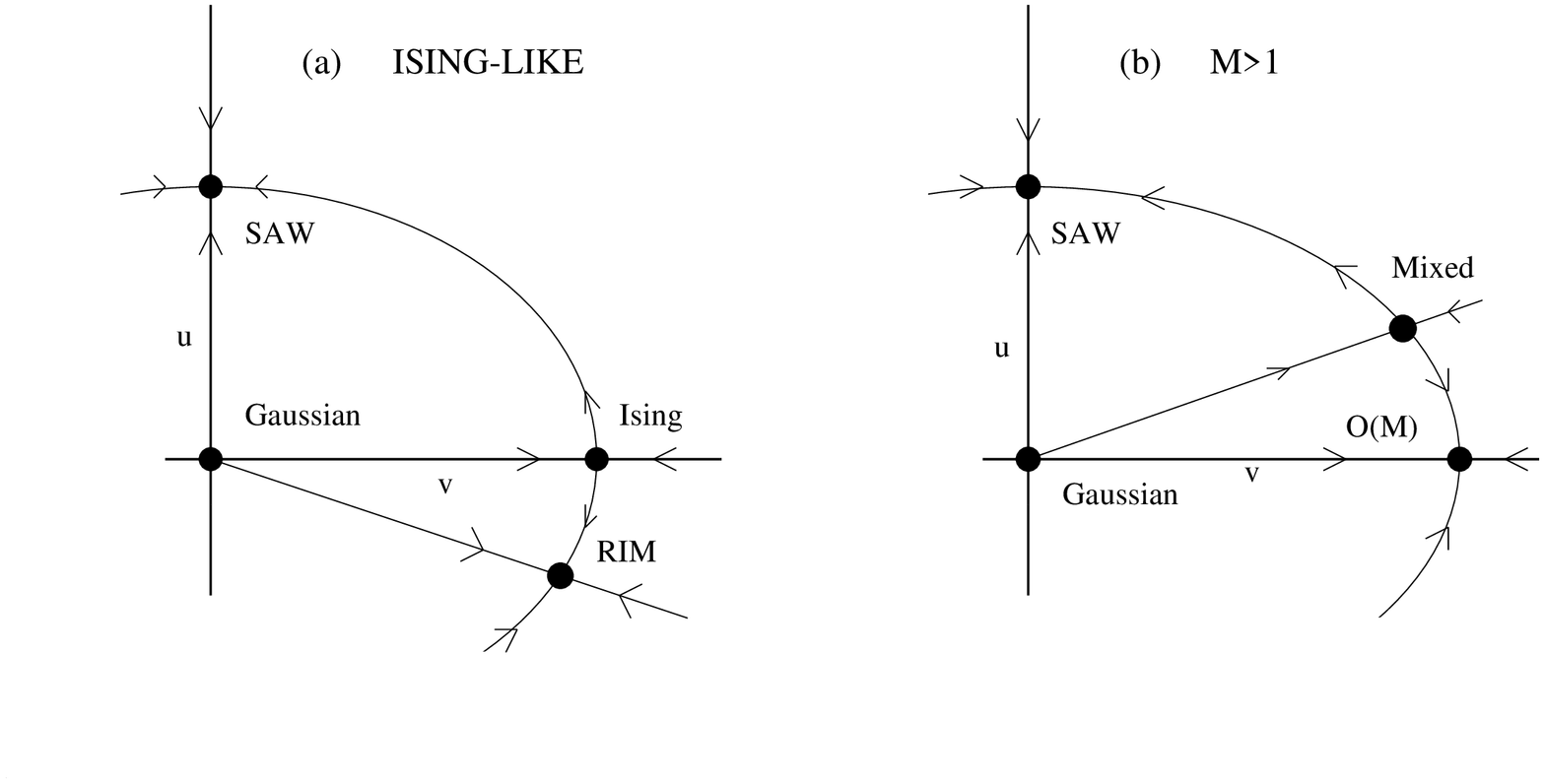}}
\vspace{-0.5cm}
\caption{
Renormalization-group flow in the coupling plane $(u,v)$ for
(a) Ising ($M=1$) and (b) $M$-component ($M>1$) systems. 
Here $N=0$.
}
\label{rgflow}
\end{figure}

The other fixed points of the Hamiltonian ${\cal H}_{M0}$ are located in 
the unphysical region $u>0$. Thus, they are not of interest for the critical 
behavior of randomly dilute spin models.
For the sake of completeness, we mention that
for $M>1$  the mixed fixed point  is in the region of positive 
$u$ and is unstable \cite{Aharony-76}.
The last fixed point is on the positive $v=0$ axis, is stable and
corresponds to the $(M N)$-vector theory for $N\rightarrow 0$. 
It is therefore in the same universality class 
of the self-avoiding walk model (SAW).
Figure~\ref{rgflow} sketches the 
flow diagram for Ising ($M=1$) and multicomponent ($M>1$) systems.

The Hamiltonian ${\cal H}_{MN}$ has been
the object of several field-theoretic studies, especially for $M=1$,
the case that describes the RIM. 
Several computations have been done in the framework
of the $\epsilon$-expansion~\cite{WF-72} and of the fixed-dimension $d=3$ 
expansion~\cite{Parisi-80}.
In these approaches, since field-theoretic perturbative expansions are 
asymptotic, the resummation of the series
is essential to obtain accurate estimates of physical quantities. For pure
systems described by the Ginzburg-Landau-Wilson Hamiltonian one exploits 
the Borel summability~\cite{bores}
of the fixed-dimension expansion (for which Borel summability is proved)
and of the $\epsilon$-expansion (for which Borel summability is conjectured), 
and the knowledge of the
large-order behavior of the series~\cite{Lipatov-77,BLZ-77}. Resummation 
procedures using these properties lead to  accurate estimates 
(see e.g. Refs. \cite{BNGM-77,LZ-77,ZJ-book,GZ-98,CPV-00}). 

Much less is known for 
the quenched disordered models described by ${\cal H}_{M0}$.
Indeed, the analytic structure of the corresponding field theory  
is much more complicated. 
The zero-dimensional model has been investigated in Ref.~\cite{BMMRY-87}.
They analyze the large-order behavior of the double expansion 
in the quartic couplings $u$ and $v$ of the free energy and 
show that the expansion in powers of $v$, 
keeping the ratio $\lambda=-u/v$ fixed, is not Borel summable. 
In Ref.~\cite{McKane-94}, it is shown that the 
non-Borel summability is
a consequence of the fact that, because of the quenched average,
there are additional singularities corresponding to the 
zeroes of the partition function $Z(\psi,g_0)$ obtained from the  
Hamiltonian (\ref{Hphi4ran}). Recently the problem has been reconsidered 
in Ref.~\cite{AMR-99}. In the same context of the zero-dimensional model,
it has been shown that a more elaborate resummation gives the 
correct determination of the
free energy from its perturbative expansion. The procedure is still 
based on a Borel summation, which is performed in two steps:
first, one resums in the coupling $v$ each 
coefficient of the series in $u$;
then, one resums the resulting series in the coupling $u$. 
There is no proof that this procedure works also in higher dimensions, 
since the method relies on the fact that the zeroes of the partition function 
stay away from the real values of $v$. This is far from obvious 
in higher-dimensional systems.

At present, the most precise field-theoretic results have been obtained 
using the fixed-dimension expansion in $d=3$. Several quantities have
been computed: 
the critical exponents \cite{Jug-83,MS-84,Shpot-89,MSS-89,%
Mayer-89,HS-92,HY-98,FHY-99,Varnashev-99,PS-99}, the equation of state
\cite{BS-92} and the hyperuniversal ratio $R^+_\xi$ \cite{BS-92,Mayer-98}.
The most precise estimates of the critical exponents for the 
RIM have been obtained from 
the analysis of the five-loop fixed-dimension expansion,
using Pad\'e-Borel-Leroy approximants~\cite{PS-99}. 
In spite of the fact that the series considered are {\em not}
Borel summable, the results for the critical exponents 
are stable: they do not depend 
on the order of the series, the details of the analysis, and, as we shall
see, are in substantial agreement with our results obtained following 
the precedure proposed in Ref.~\cite{AMR-99}.
This fact may be explained by the observation of Ref.~\cite{BMMRY-87}
that the Borel resummation applied in the standard way (i.e. at fixed $v/u$)
gives a reasonably accurate
result for small disorder if one truncates the expansion 
at an appropriate point, i.e. for not too long series.

The $MN$ model has also been extensively 
studied in the context of the $\epsilon$-expansion
\cite{HL-74,Lubensky-75,Khmelnitskii-75,AIM-76,GL-76,Aharony-76,GMM-77,JK-77,%
Shalaev-77,Newlove-83,PA-85,DG-85,Shpot-90,SAS-97}. 
The critical exponents have been computed to three loops for generic values of
$M$, $N$ \cite{DG-85} and to five loops for $M=1$ \cite{KSV-95}.
Several studies also considered the equation of state
\cite{GMM-77,Newlove-83,Shpot-90} and the two-point correlation function
\cite{GMM-77,PA-85}. In spite of these efforts,
studies based on the $\epsilon$-expansion
have been not able to go beyond a qualitative description of the physics of 
three-dimensional randomly dilute spin models. 
The $\sqrt{\epsilon}$-expansion \cite{Khmelnitskii-75} turns out not to be 
effective for a quantitative
study of the RIM (see e.g. the analysis of the five-loop series done in 
Ref. \cite{SAS-97}). A strictly related scheme is the so-called
minimal-subtraction renormalization scheme without 
$\epsilon$-expansion~\cite{SD-89}. 
The three-loop \cite{JOS-95} and four-loop \cite{FHY-98,FHY-99,FHY-00}
results are in reasonable agreement with the estimates obtained 
by other methods. At five loops, however, no random fixed point 
can be found \cite{FHY-00} using this method. This negative result has been 
interpreted as a consequence of the 
non-Borel summability of the perturbative expansion. In this case,
the four-loop series could represent the ``optimal" truncation.
We also mention that the Hamiltonian (\ref{Hphi4}) for
$M=1$ and $N\rightarrow 0$  
has been studied by the scaling-field method~\cite{NR-82}.

The randomly dilute Ising model (\ref{latticeH}) has been investigated by many 
numerical simulations (see e.g. Refs.
\cite{Landau-80,MLT-86,CS-86,BS-88,Betal-89,WS-89,WWMC-90,HF-90,%
Heuer-90,WHF-92,Heuer-93,Hennecke-93,WD-98,BFMMPR-98}).
The first simulations were apparently finding critical exponents 
depending on the spin concentration. It was later remarked 
\cite{Heuer-93,JOS-95}
that this could be simply a crossover effect: the simulations 
were not probing the critical region and were computing effective exponents
strongly biased by the corrections to scaling. Recently,
the critical exponents have been computed \cite{BFMMPR-98}
using finite-size scaling techniques. They found very strong corrections
to scaling, decaying with a rather small exponent $\omega \simeq 0.37(6)$,
--- correspondingly $\Delta=\omega\nu = 0.25(4)$ --- which is approximately a 
factor of two smaller than the corresponding pure-case
exponent. By taking into proper account the confluent corrections, 
they were able to show that the critical exponents are universal with 
respect to variations of
the spin concentration in a wide interval above the percolation point.
Their final estimates are reported in Table \ref{exponentssummary}.
 
In this paper we compute the renormalization-group functions of the generic 
$MN$ model to 
six loops in the framework of the fixed-dimension $d=3$ expansion. 
We extend the three-loop series of Ref.~\cite{Shpot-89} and the 
expansions for the cubic model ($M=1$) reported in 
Ref. \cite{PS-99} (five loops) and Ref. \cite{CPV-00} (six loops).
We will focus here on the case $N=0$ corresponding to disordered
dilute systems. Higher values of $N$ are of interest for 
several types of magnetic and structural phase transitions 
and will be discussed in a separate paper. For $M=1$ and $N\ge 2$
the six-loop series have already been analyzed in Ref. \cite{CPV-00} 
where we investigated the stability of the $O(N)$-symmetric point
in the presence of cubic interactions. We should mention that two-loop
and three-loop series for the $MN$ model in the fixed dimension expansion
for generic values of $d$ have been reported in Refs. \cite{HS-92,HY-98}.

\begin{table}[tbp]
\caption{
Best theoretical estimates of the critical exponents for the 
RIM universality class. We report results for the massive
scheme in fixed dimension $d=3$ (``$d=3$ exp."), 
the minimal subtraction scheme without $\epsilon$-expansion
(``$d=3$ MS"), and the best Monte Carlo (``MC") results.
}
\label{exponentssummary}
\begin{tabular}{cccccc}
  &
\multicolumn{1}{c}{Method}&
\multicolumn{1}{c}{$\gamma$}&
\multicolumn{1}{c}{$\nu$}&
\multicolumn{1}{c}{$\eta$}&
\multicolumn{1}{c}{$\omega$} \\
\hline
This work               & $d=3$ exp. $O(g^6)$ &
   1.330(17) & 0.678(10) & 0.030(3) & 0.25(10) \\
Ref. \protect\cite{PS-99}, 2000        & $d=3$ exp. $O(g^5)$ &
        1.325(3) & 0.671(5)  & 0.025(10) & 0.32(6) \\
Ref. \protect\cite{FHY-99,FHY-00}, 1999   & $d=3$ MS $O(g^4)$ &
    1.318   & 0.675 & 0.049  &  0.39(4)     \\
Ref. \protect\cite{BFMMPR-98}, 1998 & MC &
        1.342(10) & 0.6837(53) & 0.0374(45) & 0.37(6) \\
\end{tabular}
\end{table}

For $M=1$, $N=0$, we have performed several analyses of the perturbative 
series following the method proposed in Ref. \cite{AMR-99}. The analysis
of the $\beta$-functions for the determinaton of the fixed  point is 
particularly delicate and we have not been able to obtain a very 
robust estimate of the random fixed point. Nonetheless, we derive
quite accurate estimates of the critical exponents. Indeed,
their expansions are well behaved and largely insensitive to the 
exact position of the fixed point. Our final estimates are 
reported in Table \ref{exponentssummary}, together with 
estimates obtained by other approaches. The errors we quote are 
quite conservative and are related to the variation of the estimates
with the different analyses performed. The overall agreement is good:
the perturbative method appears to have a good predictive power,
in spite of the complicated analytic structure of the Borel transform
that does not allow the direct application of the resummation methods
used successfully in pure systems. For $M\ge 2$ and $N=0$ we have verified 
that no fixed point exists in the region $u<0$ and that the 
$O(M)$-symmetric fixed point is stable, confirming the general 
arguments given above.

The paper is organized as follows. In Sec. \ref{sec2}
we derive the perturbative 
series for the renormalization-group functions at six loops 
and discuss the singularities of the Borel transform. The results of 
the analyses are presented in Sec. \ref{sec3} and the final 
numerical values are 
reported in Table \ref{exponentssummary}.

\section{The fixed-dimension perturbative expansion of the 
three-dimensional $MN$ model.}
\label{sec2}

\subsection{Renormalization of the theory.}
\label{sec2a}

The fixed-dimension $\phi^4$ field-theoretic approach~\cite{Parisi-80} 
provides an accurate description of the
critical properties of $O(N)$-symmetric models 
in the high-temperature phase (see e.g. Ref.~\cite{ZJ-book}).
The method can also be extended to two-parameter $\phi^4$
models, such as the $MN$ model.
The idea is to perform an expansion in powers of appropriately defined 
zero-momentum quartic couplings. The theory is renormalized by introducing
a set of zero-momentum conditions for the (one-particle irreducible) 
two-point and four-point correlation functions:
\begin{equation}
\Gamma^{(2)}_{ai,bj}(p) =
  \delta_{ai,bj} Z_\phi^{-1} \left[ m^2+p^2+O(p^4)\right],
\label{ren1}  
\end{equation}
where $\delta_{ai,bj} \equiv \delta_{ab}\delta_{ij}$,
\begin{equation}
\Gamma^{(4)}_{ai,bj,ck,dl}(0) = 
Z_\phi^{-2} m \left( u S_{ai,bj,ck,dl} + v C_{ai,bj,ck,dl} \right),
\label{ren2}  
\end{equation}
and
\begin{eqnarray}
S_{ai,bj,ck,dl} &=& {1\over 3} 
\left(\delta_{ai,bj}\delta_{ck,dl} + \delta_{ai,ck}\delta_{bj,dl} + 
      \delta_{ai,dl}\delta_{bj,ck} \right),
\\
C_{ai,bj,ck,dl} &=& \delta_{ij}\delta_{ik}\delta_{il}\,{1\over 3} 
\left(\delta_{ab}\delta_{cd} + \delta_{ac}\delta_{bd} + 
      \delta_{ad}\delta_{bc} \right).
\end{eqnarray}
Eqs.~(\ref{ren1}) and (\ref{ren2})
relate the second-moment mass $m$, and the zero-momentum
quartic couplings $u$ and $v$ to the corresponding Hamiltonian parameters
$r$, $u_0$ and $v_0$:
\begin{equation}
u_0 = m u Z_u Z_\phi^{-2},\qquad\qquad
v_0 = m v Z_v Z_\phi^{-2}.
\end{equation}
In addition we define the function $Z_t$ through the relation
\begin{equation}
\Gamma^{(1,2)}_{ai,bj}(0) = \delta_{ai,bj} Z_t^{-1},
\label{ren3}
\end{equation}
where $\Gamma^{(1,2)}$ is the (one-particle irreducible)
two-point function with an insertion of $\case{1}{2}\phi^2$.

From the pertubative expansion of the correlation functions
$\Gamma^{(2)}$, $\Gamma^{(4)}$ and $\Gamma^{(1,2)}$ and 
the above relations, one derives the functions $Z_\phi(u,v)$, 
$Z_u(u,v)$, $Z_v(u,v)$, $Z_t(u,v)$ as a double expansion in $u$ and $v$.

The fixed points of the theory are given by 
the common  zeros of the $\beta$-functions
\begin{eqnarray}
\beta_u(u,v) &=& \left. m{\partial u\over \partial m}\right|_{u_0,v_0} ,\\
\beta_v(u,v) &=& \left. m{\partial v\over \partial m}\right|_{u_0,v_0} ,
\nonumber
\end{eqnarray}
calculated keeping $u_0$ and $v_0$ fixed.
The stability properties of the fixed points are controlled  by the matrix 
\begin{equation}
\Omega =\left(\matrix{\frac{\partial \beta_u(u,v)}{\partial u}
 & \frac{\partial \beta_u(u,v)}{\partial v}
 \cr\frac{\partial \beta_v(u,v)}{\partial u}
&  \frac{\partial \beta_v(u,v)}{\partial v}}\right),
\label{stability-matrix}
\end{equation}
computed at the given fixed point:
a fixed point is stable if both eigenvalues are positive.
The eigenvalues $\omega_i$ are related to
the leading scaling corrections, which vanish as
$\xi^{-\omega_i}\sim |t|^{\Delta_i}$ where $\Delta_i=\nu\omega_i$.

One also introduces the functions
\begin{eqnarray}
\eta_\phi(u,v) &=& {\partial \ln Z_\phi \over \partial \ln m}
= \beta_u {\partial \ln Z_\phi \over \partial u} +
\beta_v {\partial \ln Z_\phi \over \partial v} ,\\
\eta_t(u,v) &=& {\partial \ln Z_t \over \partial \ln m}
= \beta_u {\partial \ln Z_t \over \partial u} +
\beta_v {\partial \ln Z_t \over \partial v}.
\end{eqnarray}
Finally, the critical exponents are obtained from
\begin{eqnarray}
\eta &=& \eta_\phi(u^*,v^*),
\label{eta_fromtheseries} \\
\nu &=& \left[ 2 - \eta_\phi(u^*,v^*) + \eta_t(u^*,v^*)\right] ^{-1},
\label{nu_fromtheseries} \\
\gamma &=& \nu (2 - \eta).
\label{gamma_fromtheseries} 
\end{eqnarray}

\subsection{The perturbative series to six loops.}
\label{sec2b}

We have computed the perturbative expansion of the 
correlation functions (\ref{ren1}), (\ref{ren2}) and (\ref{ren3})
to six loops. The diagrams contributing to the two-point and 
four-point functions to six-loop order
are reported in Ref.~\cite{NMB-77}: they are approximately one thousand, 
and it is therefore necessary to handle them with a symbolic manipulation
program. For this purpose, we wrote a package in 
{\sc Mathematica} \cite{Wolfram}.
It generates the diagrams using the algorithm described in Ref.~\cite{Heap-66},
and computes the symmetry and group factors of 
each of them. We did not calculate the integrals associated to each diagram,
but we used the numerical results compiled in Ref.~\cite{NMB-77}.
Summing all contributions we determined the renormalization constants
and all renormalization-group functions.

We report our results in terms of the rescaled couplings
\begin{equation}
u \equiv  {16 \pi\over 3} \; R_{MN} \; \bar{u},\qquad\qquad
v \equiv   {16 \pi\over 3} \; R_M \;\bar{v} ,
\label{resc}
\end{equation}
where $R_K = 9/(8+K)$,
so that the $\beta$-functions associated to $\bar{u}$ and $\bar{v}$ have the 
form $\beta_{\bar{u}}(\bar{u},0) = -\bar{u} + \bar{u}^2 + O(\bar{u}^3)$ and 
$\beta_{\bar{v}}(0,\bar{v}) = -\bar{v} + \bar{v}^2 + O(\bar{v}^3)$. 

The resulting series are
\begin{eqnarray}
\beta_{\bar{u}}  =&& 
-\bar{u} + \bar{u}^2 + 
 {2(2+M)\over 8+M} \bar{u} \bar{v} 
- {4(190+41MN)\over 27(8+MN)^2} \bar{u}^3 \label{bu}\\
&&- {400(2+M)\over 27(8+MN)(8+M)} \bar{u}^2 \bar{v} - {92(2+M)\over 27(8+M)^2}\bar{u}\bar{v}^2 +
\bar{u} \sum_{i+j\geq 3} b^{(u)}_{ij} \bar{u}^i \bar{v}^j, \nonumber 
\end{eqnarray}
\begin{eqnarray}
\beta_{\bar{v}}  = && -\bar{v} + \bar{v}^2 + {12\over 8+MN} \bar{u} \bar{v} - 
{4(190+41M)\over 27(8+M)^2} \bar{v}^3 
- {16(131+25M)\over 27(8+MN)(8+M)} \bar{u}\bar{v}^2 \label{bv} \\
&&- {4(370+23MN)\over 27(8+MN)^2} \bar{u}^2 \bar{v} +
\bar{v} \sum_{i+j\geq 3} b^{(v)}_{ij} \bar{u}^i \bar{v}^j, \nonumber 
\end{eqnarray}
\begin{equation}
\eta_\phi = 
{8 (2 + MN)\over 27(8+MN)^2 } \bar{u}^2 + {16(2+M)\over 27(8+MN)(8+M) } \bar{u} \bar{v} + 
{8(2+M) \over 27(8+M)^2 } \bar{v}^2 + \sum_{i+j\geq 3} e^{(\phi)}_{ij} \bar{u}^i \bar{v}^j,
\label{etaphi}
\end{equation}
\begin{eqnarray}
\eta_t = &&-{ 2+MN\over 8+MN } \bar{u} -{2+M\over 8+M} \bar{v} +
{ 2(2+MN)\over (8+MN)^2 } \bar{u}^2 
\nonumber \\
&&+{ 4(2+M) \over (8+MN)(8+M) } \bar{u} \bar{v} 
+{ 2(2+M)\over (8+M)^2 } \bar{v}^2 
+\sum_{i+j\geq 3} e^{(t)}_{ij} \bar{u}^i \bar{v}^j.
\label{etat}
\end{eqnarray}
For $3\leq i+j\leq  6$,
The coefficients $b^{(u)}_{ij}$, $b^{(v)}_{ij}$,
$e^{(\phi)}_{ij}$ and $e^{(t)}_{ij}$ 
are reported in the Tables~\ref{betauc}, \ref{betavc},
\ref{ephi} and \ref{et} respectively.

We have performed the following checks on our calculations:
\begin{itemize}
\item[(i)]
$\beta_{\bar{u}}(\bar{u},0)$, $\eta_\phi(\bar{u},0)$ 
and $\eta_t(\bar{u},0)$ reproduce
the corresponding functions of the O($MN$)-symmetric
model~\cite{BNGM-77,AS-95};

\item[(ii)]
$\beta_{\bar{v}}(0,\bar{v})$, $\eta_\phi(0,\bar{v})$ and $\eta_t(0,\bar{v})$ reproduce
the corresponding functions of the O($M$)-symmetric model~\cite{BNGM-77,AS-95};

\item[(iii)]
For $M=1$, the functions  $\beta_{\bar{u}}$, $\beta_{\bar{v}}$,  $\eta_\phi$ 
and $\eta_t$ reproduce 
the corresponding functions of the $N$-component cubic model~\cite{CPV-00};

\item[(iv)]
The following relations hold for $N=1$:
\begin{eqnarray}
&& \beta_{\bar{u}}(u,x-u) + \beta_{\bar{v}}(u,x-u) = \beta_{\bar{v}}(0,x),\\
&&\eta_\phi(u,x-u) = \eta_\phi(0,x) ,\nonumber\\
&&\eta_t(u,x-u) = \eta_t(0,x).\nonumber
\end{eqnarray}
\end{itemize}
\begin{table}[tbp]
\squeezetable
\caption{
The coefficients $b^{(u)}_{ij}$, cf. Eq.(\ref{bu}).
}
\label{betauc}
\begin{tabular}{cl}
\multicolumn{1}{c}{$i,j$}&
\multicolumn{1}{c}{$R_{MN}^{-i} R_M^{-j} b^{(u)}_{ij}$}\\
\tableline \hline
3 , 0  &$  0.27385517 + 0.075364029\,M\,N + 0.0018504016\,{M^2}\,{N^2} $\\\hline 
2 , 1  &$  0.4516155 + 0.22580775\,M + 0.018235606\,M\,N + 
   0.0091178029\,{M^2}\,N $\\ \hline
1 , 2  &$  0.25029007 + 0.15182596\,M + 0.013340463\,{M^2} + 
   0.0017061432\,M\,N + 0.00085307161\,{M^2}\,N $\\\hline 
0 , 3  &$  0.051324521 + 0.03463704\,M + 0.00448739\,{M^2} $\\\hline\hline
4 , 0  &$  -0.27925724 - 0.091833749\,M\,N - 0.0054595646\,{M^2}\,{N^2} + 
   0.000023722893\,{M^3}\,{N^3} $\\ \hline
3 , 1  &$  -0.62922442 - 0.31461221\,M - 0.055501872\,M\,N - 
   0.027750936\,{M^2}\,N $\\
&$+ 0.00041240116\,{M^2}\,{N^2} + 0.00020620058\,{M^3}\,{N^2} $\\ \hline
2 , 2  &$  -0.55267483 - 0.36698183\,M - 0.045322209\,{M^2} - 
   0.0092541768\,M\,N $\\
&$ - 0.0036796747\,{M^2}\,N + 0.00047370687\,{M^3}\,N $\\ \hline
1 , 3  &$  -0.23360529 - 0.16634719\,M - 0.02383493\,{M^2} + 
   0.00046867235\,{M^3} $\\
&$- 0.0010494477\,M\,N - 0.00065590478\,{M^2}\,N - 0.000065590478\,{M^3}\,N $\\ \hline 
0 , 4  &$  -0.040934998 - 0.029976629\,M - 0.0046082314\,{M^2} + 
   0.000073166787\,{M^3} $\\ \hline \hline
5 , 0  &$  0.35174477 + 0.13242502\,M\,N + 0.011322026\,{M^2}\,{N^2} + 
   0.000054833719\,{M^3}\,{N^3} $\\
&$+ 8.6768933\,{{10}^{-7}}\,{M^4}\,{N^4} $\\\hline 
4 , 1  &$  1.0139338 + 0.50696692\,M + 0.12967024\,M\,N + 
   0.064835121\,{M^2}\,N + 0.00073857427\,{M^2}\,{N^2} $\\
&$+ 0.00036928714\,{M^3}\,{N^2} + 0.000021186521\,{M^3}\,{N^3} + 
   0.000010593261\,{M^4}\,{N^3} $\\ \hline
3 , 2  &$  1.2323698 + 0.85537154\,M + 0.11959332\,{M^2} + 
   0.041310322\,M\,N + 0.02290239\,{M^2}\,N $\\
&$+ 0.0011236145\,{M^3}\,N + 
   0.00017140951\,{M^2}\,{N^2} + 0.00015195691\,{M^3}\,{N^2} + 0.000033126079\,{M^4}\,{N^2} $\\ \hline
2 , 3  &$  0.80970369 + 0.61305431\,M + 0.10724129\,{M^2} + 
   0.0015700281\,{M^3} + 0.0058598808\,M\,N $\\
&$+ 0.0041006979\,{M^2}\,N + 
   0.00067267459\,{M^3}\,N + 0.000043647916\,{M^4}\,N $\\ \hline
1 , 4  &$  0.28991291 + 0.22599317\,M + 0.04307445\,{M^2} + 
   0.0013403006\,{M^3} + 0.00003112665\,{M^4} $\\
&$+ 0.0007752054\,M\,N + 
   0.0005229621\,{M^2}\,N + 0.000055080316\,{M^3}\,N - 
   6.2996919\,{{10}^{-6}}\,{M^4}\,N $\\ \hline
0 , 5  &$  0.044655379 + 0.035455913\,M + 0.0071030237\,{M^2} + 
   0.00027580942\,{M^3} + 3.1767371\,{{10}^{-6}}\,{M^4} $\\ \hline \hline
6 , 0  &$  -0.51049889 - 0.21485252\,M\,N - 0.023839375\,{M^2}\,{N^2} - 
   0.00050021682\,{M^3}\,{N^3} $\\
&$+ 2.0167763\,{{10}^{-6}}\,{M^4}\,{N^4} + 
   4.4076733\,{{10}^{-8}}\,{M^5}\,{N^5} $\\ \hline
5 , 1  &$  -1.7989389 - 0.89946945\,M - 0.30045501\,M\,N - 
   0.15022751\,{M^2}\,N - 0.007214312\,{M^2}\,{N^2} $\\
&$- 
   0.003607156\,{M^3}\,{N^2} + 0.000038644454\,{M^3}\,{N^3} + 
   0.000019322227\,{M^4}\,{N^3} + 1.3676971\,{{10}^{-6}}\,{M^4}\,{N^4} $\\
&$+ 
   6.8384853\,{{10}^{-7}}\,{M^5}\,{N^4} $\\ \hline
4 , 2  &$  -2.8025568 - 2.0077582\,M - 0.3032399\,{M^2} - 0.15673528\,M\,N - 
   0.10009354\,{M^2}\,N $\\
&$- 0.010862953\,{M^3}\,N - 
   0.00057173496\,{M^2}\,{N^2} - 0.00013491592\,{M^3}\,{N^2} + 
   0.000075475779\,{M^4}\,{N^2} $\\
&$+ 7.5116588\,{{10}^{-6}}\,{M^3}\,{N^3} + 
   9.1864196\,{{10}^{-6}}\,{M^4}\,{N^3} + 
   2.7152951\,{{10}^{-6}}\,{M^5}\,{N^3} $\\ \hline
3 , 3  &$  -2.5110555 - 2.0026222\,M - 0.40321804\,{M^2} - 
   0.014835404\,{M^3} - 0.038290982\,M\,N $\\
&$- 0.026220763\,{M^2}\,N - 
   0.0032211897\,{M^3}\,N + 0.0001582232\,{M^4}\,N - 
   0.000012833973\,{M^2}\,{N^2} $\\
&$+ 0.000015558757\,{M^3}\,{N^2} + 
   0.000020698159\,{M^4}\,{N^2} + 4.8551437\,{{10}^{-6}}\,{M^5}\,{N^2} $\\ \hline 
2 , 4  &$  -1.3671792 - 1.1287738\,M - 0.24696793\,{M^2} - 
   0.011904064\,{M^3} + 0.00014192829\,{M^4} $\\
&$- 0.0053723551\,M\,N - 
   0.0037885035\,{M^2}\,N - 0.00047438677\,{M^3}\,N + 
   0.000047179342\,{M^4}\,N $\\
&$+ 4.3956195\,{{10}^{-6}}\,{M^5}\,N $\\ \hline
1 , 5  &$  -0.42388848 - 0.35453926\,M - 0.079847022\,{M^2} - 
   0.004109703\,{M^3} + 0.00008739999\,{M^4} $\\
&$+ 
   2.4368878\,{{10}^{-6}}\,{M^5} - 0.00096920612\,M\,N - 
   0.00074246514\,{M^2}\,N - 0.00013828482\,{M^3}\,N $\\
&$- 5.8095838\,{{10}^{-6}}\,{M^4}\,N - 5.6634686\,{{10}^{-7}}\,{M^5}\,N $\\ \hline
0 , 6  &$  -0.057509877 - 0.048617439\,M - 0.011215648\,{M^2} - 
   0.00061966388\,{M^3} + 0.000011640607\,{M^4} $\\
&$+ 1.8658758\,{{10}^{-7}}\,{M^5} $\\ 
\end{tabular}
\end{table}

\begin{table}[tbp]
\squeezetable
\caption{The coefficients $b^{(v)}_{ij}$, cf. Eq.(\ref{bv}).
}
\label{betavc}
\begin{tabular}{cl}
\multicolumn{1}{c}{$i,j$}&
\multicolumn{1}{c}{$R_{MN}^{-i} R_M^{-j} b^{(v)}_{ij}$}\\
\tableline \hline
3 , 0  &$  0.64380517 + 0.05741276\,M\,N - 0.0017161966\,{M^2}\,{N^2} $\\ \hline 
2 , 1  &$  1.3928409 + 0.29248953\,M + 0.0061625366\,M\,N - 
   0.0030911252\,{M^2}\,N $\\ \hline 
1 , 2  &$  1.0440961 + 0.26681908\,M + 0.0029142164\,{M^2} $\\ \hline 
0 , 3  &$  0.27385517 + 0.075364029\,M + 0.0018504016\,{M^2} $\\\hline\hline  
4 , 0  &$  -0.76706177 - 0.089054667\,M\,N + 0.000040711369\,{M^2}\,{N^2} - 
   0.000087586118\,{M^3}\,{N^3} $\\ \hline 
3 , 1  &$  -2.2398975 - 0.49868661\,M - 0.043414868\,M\,N - 
   0.0058040064\,{M^2}\,N + 0.00021024326\,{M^2}\,{N^2} $\\
&$- 
   0.00023647795\,{M^3}\,{N^2} $\\ \hline 
2 , 2  &$  -2.5589671 - 0.75680382\,M - 0.031949535\,{M^2} + 
   0.0058629697\,M\,N + 0.0018447227\,{M^2}\,N $\\
&$- 0.00016585294\,{M^3}\,N $\\ \hline 
1 , 3  &$  -1.3553512 - 0.42919212\,M - 0.022689592\,{M^2} + 
   0.000045447676\,{M^3} $\\ \hline 
0 , 4  &$  -0.27925724 - 0.091833749\,M - 0.0054595646\,{M^2} + 
   0.000023722893\,{M^3} $\\\hline\hline 
5 , 0  &$  1.0965348 + 0.15791293\,M\,N + 0.0023584631\,{M^2}\,{N^2} - 
   0.000061471346\,{M^3}\,{N^3} $\\
&$- 5.3871247\,{{10}^{-6}}\,{M^4}\,{N^4} $\\\hline  
4 , 1  &$  4.0438017 + 0.94274678\,M + 0.1469466\,M\,N + 
   0.028781322\,{M^2}\,N - 0.0016180434\,{M^2}\,{N^2} $\\
&$- 
   0.0004537935\,{M^3}\,{N^2} + 7.2782691\,{{10}^{-7}}\,{M^3}\,{N^3} - 
   0.000020110739\,{M^4}\,{N^3} $\\ \hline 
3 , 2  &$  6.2251917 + 2.0216275\,M + 0.11770913\,{M^2} + 
   0.0079585974\,M\,N - 0.002803581\,{M^2}\,N $\\
&$- 0.0011929602\,{M^3}\,N + 
   0.00019299342\,{M^2}\,{N^2} + 0.000046931931\,{M^3}\,{N^2} - 
   0.000026294129\,{M^4}\,{N^2} $\\ \hline 
2 , 3  &$  4.9862586 + 1.7772486\,M + 0.13144724\,{M^2} - 
   0.00035323931\,{M^3} - 0.017641691\,M\,N $\\
&$- 0.0052142607\,{M^2}\,N - 
   0.00021963586\,{M^3}\,N - 0.000011811618\,{M^4}\,N $\\ \hline  
1 , 4  &$  2.0658132 + 0.75909418\,M + 0.060829135\,{M^2} + 
   0.000053192889\,{M^3} + 2.0293989\,{{10}^{-6}}\,{M^4} $\\ \hline 
0 , 5  &$  0.35174477 + 0.13242502\,M + 0.011322026\,{M^2} + 
   0.000054833719\,{M^3} + 8.6768933\,{{10}^{-7}}\,{M^4} $\\\hline \hline 
6 , 0  &$  -1.7745533 - 0.30404316\,M\,N - 0.0094338079\,{M^2}\,{N^2} + 
   0.000066993864\,{M^3}\,{N^3} $\\
&$- 6.5724895\,{{10}^{-6}}\,{M^4}\,{N^4} - 
   3.753114\,{{10}^{-7}}\,{M^5}\,{N^5} $\\ \hline 
5 , 1  &$  -7.9179198 - 1.9119099\,M - 0.4354995\,M\,N - 
   0.09835005\,{M^2}\,N + 0.0016283562\,{M^2}\,{N^2} $\\
&$+ 
   0.00057496901\,{M^3}\,{N^2} - 0.000089164587\,{M^3}\,{N^3} - 
   0.000041503636\,{M^4}\,{N^3} $\\
&$- 8.0625922\,{{10}^{-7}}\,{M^4}\,{N^4} - 
   1.7896837\,{{10}^{-6}}\,{M^5}\,{N^4} $\\ \hline 
4 , 2  &$  -15.356405 - 5.3437616\,M - 0.37337066\,{M^2} - 
   0.13516326\,M\,N - 0.031925761\,{M^2}\,N $\\
&$+ 0.00080205375\,{M^3}\,N + 
   0.00036992499\,{M^2}\,{N^2} - 0.00027756993\,{M^3}\,{N^2} - 
   0.00010718274\,{M^4}\,{N^2} $\\
&$+ 8.9627582\,{{10}^{-6}}\,{M^3}\,{N^3} - 
   1.1514477\,{{10}^{-6}}\,{M^4}\,{N^3} - 
   3.3124581\,{{10}^{-6}}\,{M^5}\,{N^3} $\\ \hline 
3 , 3  &$  -16.500282 - 6.4536377\,M - 0.60891787\,{M^2} - 
   0.0068863707\,{M^3} + 0.067945799\,M\,N $\\
&$+ 0.025945701\,{M^2}\,N + 
   0.0019347624\,{M^3}\,N - 0.00010939543\,{M^4}\,N - 
   0.00064953302\,{M^2}\,{N^2} $\\
&$- 0.00017752901\,{M^3}\,{N^2} - 
   9.1250241\,{{10}^{-6}}\,{M^4}\,{N^2} - 
   2.8529338\,{{10}^{-6}}\,{M^5}\,{N^2} $\\ \hline 
2 , 4  &$  -10.296588 - 4.1925497\,M - 0.43192146\,{M^2} - 
   0.0069161245\,{M^3} - 0.000022535528\,{M^4} $\\
&$+ 0.036155315\,M\,N + 
   0.011884073\,{M^2}\,N + 0.00065955918\,{M^3}\,N - 
   0.000016702575\,{M^4}\,N $\\
&$- 9.4492956\,{{10}^{-7}}\,{M^5}\,N $\\ \hline 
1 , 5  &$  -3.5159823 - 1.4553502\,M - 0.15565998\,{M^2} - 
   0.0028818538\,{M^3} + 2.4768276\,{{10}^{-6}}\,{M^4} $\\
&$+ 
   1.2194955\,{{10}^{-7}}\,{M^5} $\\ \hline 
0 , 6  &$  -0.51049889 - 0.21485252\,M - 0.023839375\,{M^2} - 
   0.00050021682\,{M^3} + 2.0167763\,{{10}^{-6}}\,{M^4} $\\
&$+ 
   4.4076733\,{{10}^{-8}}\,{M^5} $\\
\end{tabular}
\end{table}

\begin{table}[tbp]
\squeezetable
\caption{
The coefficients $e^{(\phi)}_{ij}$, cf. Eq.(\ref{etaphi}).
}
\label{ephi}
\begin{tabular}{cl}
\multicolumn{1}{c}{$i,j$}&
\multicolumn{1}{c}{$R_{MN}^{-i} R_M^{-j} e^{(\phi)}_{ij}$}\\
\tableline \hline
3 , 0  &$  0.00054176134 + 0.00033860084\,M\,N + 
   0.000033860084\,{M^2}\,{N^2} $\\ \hline 
2 , 1  &$  0.001625284 + 0.00081264201\,M + 0.0002031605\,M\,N + 
   0.00010158025\,{M^2}\,N $\\ \hline 
1 , 2  &$  0.001625284 + 0.0010158025\,M + 0.00010158025\,{M^2} $\\ \hline 
0 , 3  &$  0.00054176134 + 0.00033860084\,M + 0.000033860084\,{M^2} $\\\hline\hline 
4 , 0  &$  0.00099254838 + 0.00070251807\,M\,N + 
   0.0001018116\,{M^2}\,{N^2} - 6.5516886\,{{10}^{-7}}\,{M^3}\,{N^3} $\\ \hline 
3 , 1  &$  0.0039701935 + 0.0019850968\,M + 0.00082497551\,M\,N + 
   0.00041248776\,{M^2}\,N - 5.2413509\,{{10}^{-6}}\,{M^2}\,{N^2} $\\
&$- 
   2.6206755\,{{10}^{-6}}\,{M^3}\,{N^2} $\\ \hline 
2 , 2  &$  0.0059552903 + 0.0039994684\,M + 0.00051091164\,{M^2} + 
   0.00021563998\,M\,N + 0.000099957964\,{M^2}\,N $\\
&$- 
   3.9310132\,{{10}^{-6}}\,{M^3}\,N $\\ \hline 
1 , 3  &$  0.0039701935 + 0.0028100723\,M + 0.0004072464\,{M^2} - 
   2.6206755\,{{10}^{-6}}\,{M^3} $\\ \hline 
0 , 4  &$  0.00099254838 + 0.00070251807\,M + 0.0001018116\,{M^2} - 
   6.5516886\,{{10}^{-7}}\,{M^3} $\\ \hline\hline 
5 , 0  &$  -0.00036659735 - 0.0002572117\,M\,N - 
   0.000032026611\,{M^2}\,{N^2} + 2.2430702\,{{10}^{-6}}\,{M^3}\,{N^3} $\\
&$- 1.1094045\,{{10}^{-7}}\,{M^4}\,{N^4} $\\ \hline 
4 , 1  &$  -0.0018329867 - 0.00091649336\,M - 0.00036956513\,M\,N - 
   0.00018478256\,{M^2}\,N + 0.000024649511\,{M^2}\,{N^2} $\\
&$+ 
   0.000012324756\,{M^3}\,{N^2} - 1.1094045\,{{10}^{-6}}\,{M^3}\,{N^3} - 
   5.5470225\,{{10}^{-7}}\,{M^4}\,{N^3} $\\ \hline 
3 , 2  &$  -0.0036659735 - 0.0024800846\,M - 0.00032354892\,{M^2} - 
   0.000092032422\,M\,N - 7.9092305\,{{10}^{-7}}\,{M^2}\,N $\\
&$+ 
   0.000022612644\,{M^3}\,N + 4.0737345\,{{10}^{-6}}\,{M^2}\,{N^2} - 
   1.8194176\,{{10}^{-7}}\,{M^3}\,{N^2} - 
   1.1094045\,{{10}^{-6}}\,{M^4}\,{N^2} $\\ \hline 
2 , 3  &$  -0.0036659735 - 0.0026040248\,M - 0.00035159322\,{M^2} + 
   0.000016962916\,{M^3} + 0.000031907838\,M\,N $\\
&$+ 0.00003132711\,{M^2}\,N + 
   5.4677866\,{{10}^{-6}}\,{M^3}\,N - 1.1094045\,{{10}^{-6}}\,{M^4}\,N $\\ \hline  
1 , 4  &$  -0.0018329867 - 0.0012860585\,M - 0.00016013305\,{M^2} + 
   0.000011215351\,{M^3} - 5.5470225\,{{10}^{-7}}\,{M^4} $\\ \hline 
0 , 5  &$  -0.00036659735 - 0.0002572117\,M - 0.000032026611\,{M^2} + 
   2.2430702\,{{10}^{-6}}\,{M^3} - 1.1094045\,{{10}^{-7}}\,{M^4} $\\\hline\hline  
6 , 0  &$  0.00069568037 + 0.00056585941\,M\,N + 
   0.00012057302\,{M^2}\,{N^2} + 5.7466979\,{{10}^{-6}}\,{M^3}\,{N^3} $\\
&$- 
   3.8385183\,{{10}^{-8}}\,{M^4}\,{N^4} - 
   1.0441273\,{{10}^{-8}}\,{M^5}\,{N^5} $\\ \hline 
5 , 1  &$  0.0041740822 + 0.0020870411\,M + 0.0013081154\,M\,N + 
   0.00065405768\,{M^2}\,N + 0.000069380438\,{M^2}\,{N^2} $\\
&$+ 
   0.000034690219\,{M^3}\,{N^2} - 2.1003164\,{{10}^{-7}}\,{M^3}\,{N^3} - 
   1.0501582\,{{10}^{-7}}\,{M^4}\,{N^3} - 
   1.2529528\,{{10}^{-7}}\,{M^4}\,{N^4} $\\
&$- 6.264764\,{{10}^{-8}}\,{M^5}\,{N^4}\
 $\\ \hline 
4 , 2  &$  0.010435206 + 0.0074204829\,M + 0.0011014401\,{M^2} + 
   0.0010674083\,M\,N + 0.00069849884\,{M^2}\,N $\\
&$+ 0.00008239735\,{M^3}\,N + 
   8.6563947\,{{10}^{-6}}\,{M^2}\,{N^2} + 
   3.9934653\,{{10}^{-6}}\,{M^3}\,{N^2} - 
   1.6736603\,{{10}^{-7}}\,{M^4}\,{N^2} $\\
&$- 
   1.9034703\,{{10}^{-7}}\,{M^3}\,{N^3} - 
   4.0841171\,{{10}^{-7}}\,{M^4}\,{N^3} - 1.566191\,{{10}^{-7}}\,{M^5}\,{N^3}\ $\\ \hline 
3 , 3  &$  0.013913607 + 0.010984227\,M + 0.0021785508\,{M^2} + 
   0.000082419651\,{M^3} + 0.00033296155\,M\,N $\\
&$+ 0.00023139203\,{M^2}\,N + 
   0.000031719215\,{M^3}\,N - 3.6820589\,{{10}^{-7}}\,{M^4}\,N + 
   1.5175718\,{{10}^{-6}}\,{M^2}\,{N^2} $\\
&$+ 
   7.9509223\,{{10}^{-7}}\,{M^3}\,{N^2} - 
   3.9949777\,{{10}^{-7}}\,{M^4}\,{N^2} - 
   2.0882547\,{{10}^{-7}}\,{M^5}\,{N^2} $\\ \hline 
2 , 4  &$  0.010435206 + 0.0084674045\,M + 0.001790972\,{M^2} + 
   0.000082410852\,{M^3} - 3.1236483\,{{10}^{-7}}\,{M^4} $\\
&$+ 
   0.000020486706\,M\,N + 0.000017623286\,{M^2}\,N + 
   3.7896168\,{{10}^{-6}}\,{M^3}\,N - 2.6341291\,{{10}^{-7}}\,{M^4}\,N $\\
&$- 
   1.566191\,{{10}^{-7}}\,{M^5}\,N $\\ \hline 
1 , 5  &$  0.0041740822 + 0.0033951565\,M + 0.00072343812\,{M^2} + 
   0.000034480187\,{M^3} - 2.303111\,{{10}^{-7}}\,{M^4} $\\
&$- 
   6.264764\,{{10}^{-8}}\,{M^5} $\\ \hline 
0 , 6  &$  0.00069568037 + 0.00056585941\,M + 0.00012057302\,{M^2} + 
   5.7466979\,{{10}^{-6}}\,{M^3} - 3.8385183\,{{10}^{-8}}\,{M^4} $\\
&$- 
   1.0441273\,{{10}^{-8}}\,{M^5} $\\
\end{tabular}
\end{table}

\begin{table}[tbp]
\squeezetable
\caption{
The coefficients $e^{(t)}_{ij}$, cf. Eq.(\ref{etat}).
}
\label{et}
\begin{tabular}{cl}
\multicolumn{1}{c}{$i,j$}&
\multicolumn{1}{c}{$R_{MN}^{-i} R_M^{-j} e^{(t)}_{ij}$}\\
\tableline \hline
3 , 0  &$  -0.025120499 - 0.016979919\,M\,N - 0.0022098349\,{M^2}\,{N^2} $\\ \hline 
2 , 1  &$  -0.075361497 - 0.037680748\,M - 0.013259009\,M\,N - 
   0.0066295047\,{M^2}\,N $\\ \hline 
1 , 2  &$  -0.075361497 - 0.049233615\,M - 0.0057764331\,{M^2} - 
   0.0017061432\,M\,N - 0.00085307161\,{M^2}\,N $\\\hline  
0 , 3  &$  -0.025120499 - 0.016979919\,M - 0.0022098349\,{M^2} $\\\hline\hline  
4 , 0  &$  0.021460047 + 0.015690833\,M\,N + 0.0024059273\,{M^2}\,{N^2} - 
   0.000037238563\,{M^3}\,{N^3} $\\ \hline 
3 , 1  &$  0.08584019 + 0.042920095\,M + 0.019843236\,M\,N + 
   0.0099216178\,{M^2}\,N - 0.0002979085\,{M^2}\,{N^2} $\\
&$- 
   0.00014895425\,{M^3}\,{N^2} $\\ \hline 
2 , 2  &$  0.12876028 + 0.087481041\,M + 0.011550449\,{M^2} + 
   0.0066639546\,M\,N + 0.0028851146\,{M^2}\,N $\\
&$- 0.00022343138\,{M^3}\,N $\\ \hline 
1 , 3  &$  0.08584019 + 0.061713883\,M + 0.0089678045\,{M^2} - 
   0.00021454473\,{M^3} + 0.0010494477\,M\,N $\\
&$+ 0.00065590478\,{M^2}\,N + 
   0.000065590478\,{M^3}\,N $\\ \hline 
0 , 4  &$  0.021460047 + 0.015690833\,M + 0.0024059273\,{M^2} - 
   0.000037238563\,{M^3} $\\\hline \hline 
5 , 0  &$  -0.022694287 - 0.017985168\,M\,N - 0.0035835384\,{M^2}\,{N^2} - 
   0.00013566164\,{M^3}\,{N^3} $\\
&$- 1.699309\,{{10}^{-6}}\,{M^4}\,{N^4} $\\\hline  
4 , 1  &$  -0.11347143 - 0.056735717\,M - 0.033190124\,M\,N - 
   0.016595062\,{M^2}\,N - 0.0013226302\,{M^2}\,{N^2} $\\
&$- 
   0.00066131512\,{M^3}\,{N^2} - 0.00001699309\,{M^3}\,{N^3} - 
   8.496545\,{{10}^{-6}}\,{M^4}\,{N^3} $\\ \hline 
3 , 2  &$  -0.22694287 - 0.15931239\,M - 0.022920477\,{M^2} - 
   0.020539294\,M\,N - 0.012755939\,{M^2}\,N $\\
&$- 0.001243146\,{M^3}\,N - 
   0.00015896839\,{M^2}\,{N^2} - 0.00011347037\,{M^3}\,{N^2} - 
   0.00001699309\,{M^4}\,{N^2} $\\ \hline 
2 , 3  &$  -0.22694287 - 0.1746255\,M - 0.032245328\,{M^2} - 
   0.00083414751\,{M^3} - 0.0052261822\,M\,N $\\
&$- 0.0035900565\,{M^2}\,N - 
   0.00052246891\,{M^3}\,N - 0.00001699309\,{M^4}\,N $\\ \hline 
1 , 4  &$  -0.11347143 - 0.089150636\,M - 0.01739473\,{M^2} - 
   0.00062322789\,{M^3} - 0.000014796237\,{M^4} $\\
&$- 0.0007752054\,M\,N - 
   0.0005229621\,{M^2}\,N - 0.000055080316\,{M^3}\,N + 
   6.2996919\,{{10}^{-6}}\,{M^4}\,N $\\ \hline 
0 , 5  &$  -0.022694287 - 0.017985168\,M - 0.0035835384\,{M^2} - 
   0.00013566164\,{M^3} - 1.699309\,{{10}^{-6}}\,{M^4} $\\ \hline \hline
6 , 0  &$  0.029450619 + 0.024874579\,M\,N + 0.005728397\,{M^2}\,{N^2} + 
   0.00031557863\,{M^3}\,{N^3} $\\
&$- 5.858689\,{{10}^{-6}}\,{M^4}\,{N^4} - 
   1.0373506\,{{10}^{-7}}\,{M^5}\,{N^5} $\\ \hline 
5 , 1  &$  0.17670371 + 0.088351856\,M + 0.060895618\,M\,N + 
   0.030447809\,{M^2}\,N + 0.0039225729\,{M^2}\,{N^2} $\\
&$+ 
   0.0019612864\,{M^3}\,{N^2} - 0.000067814627\,{M^3}\,{N^3} - 
   0.000033907313\,{M^4}\,{N^3} - 1.2448208\,{{10}^{-6}}\,{M^4}\,{N^4} $\\
&$- 
   6.2241039\,{{10}^{-7}}\,{M^5}\,{N^4} $\\ \hline 
4 , 2  &$  0.44175928 + 0.31708477\,M + 0.048102565\,{M^2} + 
   0.056033914\,M\,N + 0.037304669\,{M^2}\,N $\\
&$+ 0.0046438563\,{M^3}\,N + 
   0.00051871955\,{M^2}\,{N^2} + 0.000095257657\,{M^3}\,{N^2} - 
   0.00008205106\,{M^4}\,{N^2} $\\
&$- 5.434447\,{{10}^{-6}}\,{M^3}\,{N^3} - 
   5.8292754\,{{10}^{-6}}\,{M^4}\,{N^3} - 1.556026\,{{10}^{-6}}\,{M^5}\,{N^3}\
 $\\ \hline 
3 , 3  &$  0.58901238 + 0.47310641\,M + 0.097561547\,{M^2} + 
   0.0041307177\,{M^3} + 0.024385168\,M\,N $\\
&$+ 0.017002488\,{M^2}\,N + 
   0.002190569\,{M^3}\,N - 0.00010719141\,{M^4}\,N + 
   3.9039471\,{{10}^{-6}}\,{M^2}\,{N^2} $\\
&$- 
   9.7139715\,{{10}^{-6}}\,{M^3}\,{N^2} - 
   9.9823751\,{{10}^{-6}}\,{M^4}\,{N^2} - 
   2.0747013\,{{10}^{-6}}\,{M^5}\,{N^2} $\\ \hline 
2 , 4  &$  0.44175928 + 0.36731802\,M + 0.081706647\,{M^2} + 
   0.0041012326\,{M^3} - 0.000071248084\,{M^4} $\\
&$+ 0.0058006649\,M\,N + 
   0.0042193071\,{M^2}\,N + 0.00063244692\,{M^3}\,N - 
   0.000016632252\,{M^4}\,N $\\
&$- 1.556026\,{{10}^{-6}}\,{M^5}\,N $\\ \hline 
1 , 5  &$  0.17670371 + 0.14827827\,M + 0.033627917\,{M^2} + 
   0.001755187\,{M^3} - 0.000040961717\,{M^4} $\\
&$- 
   1.1887572\,{{10}^{-6}}\,{M^5} + 0.00096920611\,M\,N + 
   0.00074246513\,{M^2}\,N + 0.00013828482\,{M^3}\,N $\\
&$+ 
   5.809583\,{{10}^{-6}}\,{M^4}\,N + 5.6634686\,{{10}^{-7}}\,{M^5}\,N $\\\hline  
0 , 6  &$  0.029450619 + 0.024874579\,M + 0.005728397\,{M^2} + 
   0.00031557863\,{M^3} - 5.858689\,{{10}^{-6}}\,{M^4} $\\
&$- 
   1.0373506\,{{10}^{-7}}\,{M^5} $\\
\end{tabular}
\end{table}

\subsection{Borel summability and resummation of the series.}
\label{sec2c}
 
Since field-theoretic  perturbative expansions are asymptotic, 
the resummation of the series is essential
to obtain accurate estimates of the physical quantities.

In the case of the O($N$)-symmetric $\phi^4$ theory the expansion 
is performed in powers of the zero-momentum four-point coupling $g$.
The large-order behavior of the series $S(g) = \sum s_k g^k$
of any quantity is related to the singularity $g_b$ of the Borel transform
closest to the origin. Indeed, for large $k$, the coefficient $s_k$
behaves as 
\begin{equation}
s_k \sim k! \,(-a)^{k}\, k^b \,\left[ 1 + O(k^{-1})\right] \qquad\qquad
{\rm with}\qquad a = - 1/g_b.
\label{lobh}
\end{equation}
The value of $g_b$ depends only on the Hamiltonian,
while the exponent $b$ depends on which Green's function is considered.
The value of $g_b$ can be obtained from a steepest-descent calculation in which
the relevant saddle point is a finite-energy solution (instanton)
of the classical field equations with negative 
coupling~\cite{Lipatov-77,BLZ-77}.
Since the Borel transform is singular for $g=g_b$, its expansion in powers 
of $g$ converges only for $|g|< |g_b|$.
An analytic extension 
can be obtained by a conformal mapping~\cite{LZ-77}, such as
\begin{equation}
y(g) = {\sqrt{1 - g/g_b} - 1\over \sqrt{1 - g/g_b} + 1 }.
\label{cm}
\end{equation}
In this way the Borel transform becomes a series in powers of $y(g)$
that converges for all positive values of $g$ provided that all 
singularities of the Borel transform are on the real negative 
axis~\cite{LZ-77}.  
In this case one obtains a convergent sequence of approximations 
for the original quantity.
For the O($N$)-symmetric theory accurate estimates 
(see e.g. Ref.~\cite{GZ-98}) have been obtained resumming
the available series: the $\beta$-function~\cite{BNGM-77} is known
up to six loops, while the functions $\eta_\phi$ and $\eta_t$ 
are known to seven loops~\cite{MN-91}.

The large-order behavior of the perturbative expansions 
in the $MN$ model can be studied by employing
the same methods used in the standard $\phi^4$ 
theory~\cite{PV_inprep}. We may 
consider the series in $\bar{u}$ and $\bar{v}$ at 
fixed ratio $z \equiv  \bar{v}/\bar{u}$. 
The large-order behavior of the resulting expansion
in powers of $\bar{u}$ is determined by 
the singularity of the Borel transform that is
closest to the origin, $\bar{u}_b(z)$, given by 
\begin{eqnarray}
{1\over \bar{u}_b(z)} &= - a \left( R_{MN} + R_M z\right)
\qquad\qquad &{\rm for} \qquad z >0 \qquad {\rm and} \qquad 
                 z < - {2 N\over N+1} {R_{MN}\over R_M},
\nonumber \\
{1\over \bar{u}_b(z)} &= - a \left( R_{MN} + {1\over N} R_{M}z \right)
\qquad\qquad &{\rm for} \qquad\qquad - {2 N\over N+1} {R_{MN}\over R_M} < z <0.
\label{bsing} 
\end{eqnarray}
where
\begin{equation}
a = 0.14777422...,\qquad\qquad R_{K} = {9\over 8+K}.
\label{a-and-RK}
\end{equation} 
Using Eq. (\ref{bsing}) and the conformal mapping (\ref{cm}), one can
resum the perturbative series in $\ub$ at fixed $z$. This method
has been applied in Ref.~\cite{CPV-00} to the analysis of 
the renormalization-group functions  of the three-dimensional cubic model.

The result (\ref{bsing}) has been obtained for integer $M,N\ge 1$. 
For $N=0$, one may think that the correct behaviour is obtained 
by analytic continuation of (\ref{bsing}), i.e. 
\begin{equation}
{1\over \bar{u}_b(z)} = - a \left( \case{9}{8} + R_M z\right),
\label{bsingN0}
\end{equation}
for all $z$. However, this is not correct.
Indeed, as explicitly shown in Refs.~\cite{BMMRY-87,McKane-94}
in the context of the zero-dimensional random Ising model, 
there is an additional contribution to 
the large-order behavior of the series in $u$ at fixed $\lambda\equiv -u/v$, 
which makes the series non-Borel summable, giving rise 
to singularities of the Borel transform on 
the positive real axis. They are due to the zeroes of the partition
function at fixed disorder.
We have no reason  to believe that similar non-Borel summable 
contributions are not present
in higher dimensions. It is likely that the same phenomenon
occurs even in three dimensions. As a consequence, 
a summation procedure based on  Eq. (\ref{bsingN0})
and a conformal mapping of the type (\ref{cm}) would not  
lead to a sequence of approximations converging to the correct 
result~\cite{BMMRY-87}.

Fortunately, this is not the end of the story.
As shown recently in Ref.~\cite{AMR-99}, at least in zero dimensions,
one can still resum the perturbative series.
Indeed the zero-dimensional free energy can be obtained from its perturbative
expansion if one applies a more elaborated procedure which is still based
on a Borel summation. 
Let us write the double expansion of the free energy $f(u,v)$ in 
powers of $u,v$ as
\begin{eqnarray}
f(u,v) &=& \sum_{n=0}^\infty c_n(v) u^n, 
\label{fuv} \\
c_n(v) &\equiv&  \sum_{k=0}^\infty c_{nk} v^k,
\label{cnv}
\end{eqnarray}
The main result of Ref.~\cite{AMR-99} is that the expansions of the 
coefficients (\ref{cnv}) and the resulting series at fixed $v$, 
Eq. (\ref{fuv}), are Borel summable. Using this result, a resummation
of the free energy is obtained in two steps. First, 
one resums the coefficient $c_n(v)$; then, using the computed 
coefficients, one resums the series in $u$.
The resummation of Eq. (\ref{cnv}) can be performed using the 
Pad\'e-Borel-Leroy method, as suggested in Ref.~\cite{AMR-99}.
However, also the conformal method can be used, since the large-order
behavior  is known exactly. Indeed,
\begin{equation}
c_n(v) \propto \left. {\partial^n f(u,v)\over \partial u^n}\right|_{u=0}.
\end{equation}
Thus, $c_n(v)$ can be related to zero-momentum correlation functions
in the theory with $u=0$, which is the standard $M$-vector model.
Therefore, one can use the well-known results for the 
large-order behavior of the perturbative series 
in the O($M$)-symmetric theory~\cite{Lipatov-77,BLZ-77}.

\section{Analysis of the six-loop expansion for $N=0$} \label{sec3}
 
\subsection{The random Ising model}

As we said in the Introduction, the random Ising model corresponds 
to $M=1$ and $N=0$. There are two relevant fixed points, the 
Ising and the random point, see Fig. \ref{rgflow}.
In Ref. \cite{CPV-00} we already discussed the stability
of the Ising point. We found that this fixed point is unstable
since the stability matrix has a negative eigenvalue
$\omega = - 0.177(6)$, in good agreement with the general argument
predicting $\omega = -\alpha_I/\nu_I = - 0.1745(12)$. 
We will now investigate the random fixed point, which is stable
and determines therefore the critical behaviour of the RIM.

In order to study the critical properties of the random fixed point, 
we used several different resummation procedures, according to the
discussion of the previous Section.
Following Ref. \cite{AMR-99}, for each quantity we consider,
we must perform first a resummation of 
the series in $v$, see Eq. (\ref{cnv}). This may be done in two different 
ways. We can either use the Pad\'e-Borel method, or 
perform a conformal mapping of the Borel-transformed series, using 
the known value of the singularity of the Borel transform.
Explicitly, let us consider a $p$-loop series in $u$ and $v$ of the form
\begin{equation}
    \sum_{n=0}^p \sum_{k=0}^{p-n} c_{nk} u^n v^k.
\label{original}
\end{equation}
In the first method,
for each $0\le n\le p$, we choose a real number $b_n$ and a positive 
integer $r_n$ such that $r_n \le p-n$; then, we consider 
\begin{equation}
R_1(c_n)(p;b_n,r_n;v) = 
 \, \int_0^\infty dt\, e^{-t} t^{b_n} \,
  \left[\sum_{i=0}^{p-n-r_n} B_i (tv)^i\right]\,
  \left[1 + \sum_{i=1}^{r_n} C_i (tv)^i\right]^{-1} .
\label{eq3.2}
\end{equation}
The coefficients $B_i$ and $C_i$ are fixed so that 
$R_1(c_n)(p;b_n,r_n;v) = \sum_{k=0}^{p-n} c_{nk} v^k + O(v^{p-n+1})$. 
Here we are resumming the Borel transform of each coefficient 
of the series in $u$ by means of a Pad\'e approximant $[p-n-r_n/r_n]$.
Eq. (\ref{eq3.2}) is well defined as long as the integrand is regular for all
positive values of $t$. However, for some values of the parameters, 
the Pad\'e approximant has poles on the positive real axis 
--- we will call these cases {\em defective} --- so that the integral 
does not exist. These values of $b_n$ and $r_n$ must of course be 
discarded.

The second method uses the large-order behaviour of the series 
and a conformal mapping \cite{LZ-77,ZJ-book}. 
In this case, for each $0\le n\le p$, we choose two real numbers 
$b_n$ and $\alpha_n$ and consider 
\begin{equation}
R_2(c_n)(p;b_n,r_n;v) =\,
 \sum_{k=0}^{p-n} B_k \int_0^\infty dt\, e^{-t} t^{b_n} 
   {(y(vt))^k\over (1-y(vt))^{\alpha_n}},
\end{equation}
where 
\begin{equation}
y(t) = {\sqrt{|\overline{g}_I| + t} - \sqrt{|\overline{g}_I|} \over 
        \sqrt{|\overline{g}_I| + t} + \sqrt{|\overline{g}_I|}},
\end{equation}
and $\overline{g}_I = - 1/a$ is the singularity of the Borel transform
for the pure Ising model. 
The numerical value of $a$ is given in Eq. (\ref{a-and-RK}).
Using these two methods we obtain two different partial resummations
of the original series (\ref{original}):
\begin{eqnarray}
\sum_{n=0}^p R_1(c_n)(p;b_n,r_n;v) u^n, \label{R1} \\
\sum_{n=0}^p R_2(c_n)(p;b_n,\alpha_n;v) u^n. \label{R2}
\end{eqnarray}
Nothing is known on the asymptotic behaviour of these series, and we will
thus use the Pad\'e-Borel method. Starting from Eq. (\ref{R1}) we will 
thus consider 
\begin{equation}
E_1(c)(q,p;b_u,r_u;\{b_n\},\{r_n\})\equiv 
\int_0^\infty dt\, e^{-t} t^{b_u}\; 
\left[ \sum_{i=0}^{q-r_u} B_i(v) (tu)^i\right]\,
\left[1 +  \sum_{i=1}^{r_u} C_i(v) (tu)^i\right]^{-1}.
\label{E1}
\end{equation}
The coefficients $B_i(v)$ and $C_i(v)$ are fixed so that 
$E_1(c)(q,p;b_u,r_u;\{b_n\},\{r_n\})$ coincides with the expansion
(\ref{R1}) up to terms of order $O(u^{q+1})$. Note that we have introduced here
three additional parameters: $b_u$, the power appearing in the 
Borel transform, $r_u$ that fixes the order of the Pad\'e approximant,
and $q$ which indicates the number of terms that are resummed, and that, 
in the following, will always satisfy $q\le p-1$. Analogously, starting from 
Eq. (\ref{R2}), we define 
$E_2(c)(q,p;b_u,r_u;\{b_n\},\{\alpha_n\})$.
We will call the first method the ``double Pad\'e-Borel" method, while 
the second will be named the "conformal Pad\'e-Borel" method.

Let us now apply these methods to the computation of the fixed point
$(\ub^*,\vb^*)$. In this case, we resum the $\beta$-functions
$\beta_u/u$ and $\beta_v/v$ and then look for a common zero with $u<0$. 
We consider first the resummation $E_1$, Eq. (\ref{E1}). 
A detailed analysis
shows that the coefficients $R_1$ can only be defined for $r_n = 1$. 
We have also tried $r_0=2$ and $r_1=2$, but the resulting 
Pad\'e approximants turned out defective. Therefore, we have fixed 
$r_n=1$ for all $0\le n\le 5$. We must also fix the parameters 
$\{b_n\}$. It is impossible to vary all of them independently,
since there are too many combinations. For this reason, we have taken
all $b_n$ equal, i.e. we have set $b_n=b_v$ for all $n$. 
Finally, we have only considered the case $q=p-1$. 
Therefore, the analysis is based on the approximants 
\begin{equation}
\widehat{E}_1(\cdot)(p;b_u,r_u;b_v) =  
  E_1(\cdot)(p-1,p;b_u,r_u;\{b_n=b_v\},\{r_n=1\}).
\label{def-approximantsE1}
\end{equation}
Estimates of the fixed point $(\ub^*,\vb^*)$ have been obtained 
by solving the equations
\begin{equation}
\widehat{E}_1(\beta_\ub/\ub) (p;b_u,r_u;b_v) = 0, \qquad \qquad
\widehat{E}_1(\beta_\vb/\vb) (p;b_u,r_u;b_v) = 0.
\end{equation}
We have used $r_u=1,2$, $p=4,5,6$ and we have varied $b_u$ and $b_v$
between 0 and 20.
As usual in these procedures, we must determine the optimal values of the 
parameters $b_u$ and $b_v$. This is usually accomplished by looking for 
values of $b_u$ and $b_v$ such that the estimates are essentially independent
of the order $p$ of the series. In the present case, we have not been able to 
find any such pair. Indeed, the five-loop results ($p=5$) are systematically
higher than those obtained with $p=4$ and $p=6$. For instance, if we average 
all estimates with $0\le b_u,b_v\le 5$ we obtain 
\begin{eqnarray}
\ub^* &=& -0.66(1)\; (p=4),\qquad\qquad -0.78(2)\; (p=5), 
    \qquad\qquad -0.63(3) \; (p=6); \\
\vb^* &=& 2.235(3)\; (p=4),\qquad\qquad 2.273(4)\; (p=5),
    \qquad\qquad 2.250(23) \; (p=6).
\end{eqnarray}
The uncertainties quoted here are the standard deviations of the estimates in 
the quoted interval and show that the dependence on $b_u$, $b_v$, and 
$r_u$ is very small compared to the variation of the results with $p$. 
Increasing $b_u,b_v$ does not help, since the five-loop result is
largely insensitive to variations of the parameters, while 
for $p=4$ and $p=6$ $|\ub^*|$ and 
$\vb^*$ decrease with increasing $b_u$ and $b_v$. 
It is difficult to obtain a final estimate from these results. We quote 
\begin{equation}
\ub^* = -0.68(10), \qquad\qquad \vb^* = 2.25(2),
\end{equation}
that includes all estimates reported above. 

The instability of the results reported above with $p$ seems to indicate 
that some of the hypotheses underlying the choice of the parameters is probably
incorrect. One may suspect that choosing all $b_n$ equal does not 
allow a correct resummation of the coefficients, and that 
$\beta_\ub$ and $\beta_\vb$ need different choices of the parameters.
We have therefore tried a second strategy. First, for each $\beta$-function,
we have carefully analyzed each coefficient of the series in $u$, trying to 
find an optimal value of the parameter $b_n$ --- $r_n$ was fixed in all 
cases equal to 1 --- by requiring the stability of the 
estimates of the coefficient with respect to a change of the order of 
the series. However, only for the first two
coefficients we were able to identify a stable region,
so that we could not apply this method. On the other hand, as we shall see, 
this method works very well for the resummations of the coefficients
that use the conformal mapping.

Let us now discuss the conformal Pad\'e-Borel method.
As before, we have tried two different strategies. In the first case
we have set all $b_n$ equal to $b_v$ and all $\alpha_n$ equal to 
$\alpha_v$, we have used the same parameters for the two
$\beta$-functions, and we have looked for optimal values of 
$b_u$, $b_v$ and $\alpha_v$, setting $r_u=1,2$. 
While before, for each $p$, the estimates were stable, in this case
the fluctuations for each fixed $p$ are very large, and no estimate 
can be obtained.

Then, we applied the second strategy. We 
analyzed carefully each coefficient of the series in $u$, finding
optimal values $b_{n,\rm opt}$ and $\alpha_{n,\rm opt}$ for each
$n$ and $\beta$-function. Of course, the required stability analysis 
can only be performed if the series is long enough,
and thus we have always taken $q\le 4$. 
Therefore, we consider
\begin{equation}
\widehat{E}_2(\cdot)(q,p;b_u,r_u;\delta_b,\delta_\alpha) = 
  E_2(\cdot)(q,p;b_u,r_u;\{b_{n,\rm opt} + \delta_b\},
                         \{\alpha_{n,\rm opt} + \delta_\alpha\}),
\label{def-approximantsE2}
\end{equation}
where $\delta_\alpha$ and $\delta_b$ are ($n$-independent) numbers 
which allow us to vary $b_n$ and $\alpha_n$ around the optimal 
values. Estimates of the fixed point are obtained from 
\begin{equation}
 \widehat{E}_2(\beta_\ub/\ub)(q,p;b_u,r_{1,u};\delta_b,\delta_\alpha) = 0,
\qquad\qquad 
 \widehat{E}_2(\beta_\vb/\vb)(q,p;b_u,r_{2,u};\delta_b,\delta_\alpha) = 0.
\end{equation}
The first problem which must be addressed is the value of the parameters
$r_{1,u}$ and $r_{2,u}$. For $\beta_\ub/\ub$ we find that the Pad\'e
approximants are always defective for $r_{1,u}=1,2$; 
they are well behaved only for 
$r_{1,u}=3$ and $q=4$. Since the resummed series in $u$ has coefficients that 
are quite small, we decided to use also $r_{1,u}=0$, which corresponds to 
a direct summation of the series in $u$, without any Pad\'e-Borel 
transformation.
For $\beta_\vb/\vb$ we did not observe a regular pattern 
for the defective Pad\'e's and we have used $r_{2,u}=1,2$, 
discarding all defective cases.
\begin{table}[tbp]
\caption{
Estimates of $(\ub^*,\vb^*)$ obtained using the conformal Pad\'e-Borel method.
The results 
are averages over $0\le b_u\le 20$, $-3\le \delta_b\le 3$,
$-1\le \delta_\alpha\le 1$, $r_{2,u} = 1,2$. ``Def" is the percentage of 
defective Pad\'e approximants in each analysis.
}
\label{ustarvstar}
\begin{tabular}{llrrrr}
\multicolumn{1}{c}{$q$} &
\multicolumn{1}{c}{$r_{1,u}$} & 
\multicolumn{2}{c}{$p=5$} &
\multicolumn{2}{c}{$p=6$} \\
 & & 
\multicolumn{1}{c}{($\ub^*$, $\vb^*$)} &  Def & 
\multicolumn{1}{c}{($\ub^*$, $\vb^*$)} & Def \\
\tableline \hline
3 & 0 & [$-$0.615(27), 2.175(43)] & 80\% & [$-$0.641(7), 2.199(10)] & 78\% \\
4 & 0 & [$-$0.618(22), 2.190(35)] & 74\% & [$-$0.630(9), 2.194(16)] & 61\% \\
4 & 3 & [$-$0.619(20), 2.191(33)] & 71\% & [$-$0.632(8), 2.196(15)] & 61\% \\
\end{tabular}
\end{table}
The results, for chosen values of $p$, $q$, and $r_{1,u}$ are reported in 
Table \ref{ustarvstar}. The quoted uncertainty is the standard deviation
of the results when $-3\le \delta_b\le 3$ and 
$-1\le \delta_\alpha\le 1$. This choice is completely arbitrary,
but in similar analyses of different models we found that 
varying $\alpha$ by $\pm1$ and $b$ by $\pm3$ provides a reasonable 
estimate of the error. Notice that we have not optimized $b_u$, but
we have averaged over all values between 0 and 20, since the dependence on
this parameter is extremely small.
The results are stable, giving a final
estimate (average of the results with $p=6$, $q=4$)
\begin{equation}
\ub^* = -0.631(16), \qquad\qquad \vb^* = 2.195(20).
\label{stimeustarvstar_2}
\end{equation}
The error bars have been chosen in such a way to include all central values for
$p=5$ and $p=6$. 
It should be noted that, even if our
results are quite stable with respect to changes of the parameters,
most of the approximants do not contribute since they are defective.
For these reasons, in the following we will always
carefully check the dependence of the estimates on the value of the 
fixed point, considering also values of $(\ub^*,\vb^*)$ 
that are well outside the confidence intervals of 
Eq. (\ref{stimeustarvstar_2}).

We can compare our results for the fixed point with previous determinations.
Ref.~\cite{MSS-89} reports $(\ub^*,\vb^*) = (-0.667,2.244)$ obtained from the 
Chisholm-Borel analysis of the four-loop series. The same expansion was also 
analyzed by Varnashev \cite{Varnashev-99} obtaining
$(\ub^*,\vb^*) = (-0.582(85),2.230(83))$ and 
$(\ub^*,\vb^*) = (-0.625(60),2.187(56))$ using different sets 
of Pad\'e approximants. The $\epsilon$-algorithm by Wynn 
with a Mittag-Leffler transform was used in Ref. \cite{Mayer-89}
finding $(\ub^*,\vb^*) = (-0.587,2.178)$. From the analysis of the 
five-loop series Pakhnin and Sokolov
\cite{PS-99} obtain 
$(\ub^*,\vb^*) = (-0.711(12),2.261(18))$. While the four-loop results 
are in good agreement with our estimates, the five-loop
estimate differs significantly, a fact that may indicate that the claim of 
Ref. \cite{PS-99} that the error on their estimates 
is approximately 1--2\% is rather optimistic. Note also 
that the five-loop result is quite different from the previous 
four-loop estimates.

We have also tried to determine the eigenvalues of the stability 
matrix $\Omega$, cf. Eq. (\ref{stability-matrix}), that controls the 
subleading corrections in the model. We used both a double
Pad\'e-Borel transformation and the conformal-Pad\'e-Borel method. 
In the first case we obtain estimates that vary strongly with the order, and,
as it happened for the position of the fixed point, it is impossible to obtain 
results that are insensitive to the order $p$. 
For $r_u=1$, discarding the cases in which the computed eigenvalues are 
complex, we obtain for the smallest eigenvalue $\omega$:
\begin{equation}
\omega = 0.16(2)\; (p=4), \qquad\qquad
           0.21(3)\; (p=5), \qquad\qquad
           0.16(3)\; (p=6).
\label{stimeomega2_1}
\end{equation}
We have included in the error the dependence on the position of the 
fixed point. These estimates have been obtained setting 
$r_n=1$ and averaging over $b_u$ and $b_v$ varying between 0 and 10. We have 
not tried to optimize the choice of these parameters, since the estimates 
show only a small dependence on them.
We have also considered $r_u=2$. In this case a large fraction of the 
approximants is defective (for $p=4$ they are all defective). We obtain
\begin{equation}
\omega = 0.29(4) \; (p=5), \qquad\qquad
           0.33(5) \; (p=6).
\label{stimeomega2_2}
\end{equation}
The quite large discrepancy between the estimates (\ref{stimeomega2_1})
and (\ref{stimeomega2_2}) clearly indicates that the analysis is not 
very robust. A conservative final estimate is 
\begin{equation} 
  \omega = 0.25(10),
\end{equation}
that includes the previous results. 

We have also tried the conformal-Pad\'e-Borel method, optimizing
separately each coefficient . However,
several problem appeared immediately. First, we could not 
perform a Pad\'e-Borel resummation of the series in $u$ of the 
elements of the stability matrix. Indeed, in all cases, 
some Pad\'e approximant was defective. As in the determination of the 
fixed-point position, 
we tried to resum the series in $u$  without any additional
transformation. For $p=4$ this gives reasonable results, and we can estimate
$\omega = 0.29(9)$. However, for $p=5,6$ all eigenvalues we find are
complex, and as such must be discarded. 

The fact that the series appearing in the stability matrix generate always
defective Pad\'e approximants may indicate that the series in $u$ 
are {\em not} Borel summable. In this case, one expects that the 
estimates converge towards the correct value up to a certain number of 
loops. Increasing further the length of the series, worsens the final results.
If indeed the expansion is not Borel summable, the previous results 
seem to indicate that for the subleading exponent $\omega$ 
the best results are obtained at four loops. 

Let us now compute the critical exponents. 
As before, we tried several different methods. 
A first estimate was determined using the double Pad\'e-Borel method.
Each exponent was computed from the approximants
$\widehat{E}_1(\epsilon)(p;b_u,r_u;b_v)$ defined in Eq. 
(\ref{def-approximantsE1}).
For $\gamma$ and $\nu$, the series $1/\gamma$ and $1/\nu$ are more stable
and thus the final estimates are obtained from their analysis. For $\eta$,
if we write $\eta = \sum \eta_n(v) u^n$, then $\eta_0(v)\sim v^2$ and 
$\eta_1(v) \sim v$. In this case, for $n=0$ we resummed the series
$\eta_0(v)/v^2$, while for $n=1$ we considered $\eta_1(v)/v$. 
The results we obtain are very stable, even if we do not optimize
the parameters $b_u$ and $b_v$.
\begin{table}[tbp]
\caption{Estimates of the critical exponents for the RIM universality
class obtained using a double Pad\'e-Borel resummation. The estimates 
correspond to the following choices of the parameters:
$0\le b_u,b_v\le 20$, $r_u=1,2$. The subscript indicates the percentage
of cases in which some Pad\'e approximant appearing in the resummation
procedure was defective.}
\label{stimeexp_1}
\begin{tabular}{llll}
\multicolumn{1}{c}{$(\ub^*,\vb^*)$} &
\multicolumn{1}{c}{$\nu$}  & 
\multicolumn{1}{c}{$\gamma$}  & 
\multicolumn{1}{c}{$\eta$} \\
\hline
\multicolumn{4}{c}{$p=4$} \\
\hline
$(-0.631,2.195)$  & $0.6706(3)_{49\%}$ & $1.3222(4)_{50\%}$ & \\
$(-0.640,2.230)$  & $0.6740(3)_{49\%}$ & $1.3282(5)_{50\%}$ & \\
$(-0.680,2.240)$  & $0.6725(2)_{49\%}$ & $1.3283(5)_{50\%}$ & \\
\hline
\multicolumn{4}{c}{$p=5$} \\
\hline
$(-0.631,2.195)$  & $0.6740(4)_{0\%}$ & $1.3270(7)_{0\%}$ &  
                    $0.03081(3)_{0\%}$  \\
$(-0.745,2.275)$  & $0.6739(4)_{0\%}$ & $1.3291(7)_{0\%}$ &  
                    $0.02738(2)_{0\%}$  \\
$(-0.800,2.270)$  & $0.6687(3)_{0\%}$ & $1.3213(7)_{0\%}$ &  
                    $0.02387(2)_{0\%}$  \\
\hline
\multicolumn{4}{c}{$p=6$} \\
\hline
$(-0.631,2.195)$  & $0.6677(4)_{0\%}$ & $1.3130(8)_{0\%}$ &  
                    $0.03272(2)_{50\%}$  \\
$(-0.570,2.210)$  & $0.6745(6)_{0\%}$ & $1.3225(11)_{0\%}$&  
                    $0.03810(4)_{50\%}$  \\
$(-0.700,2.290)$  & $0.6727(4)_{0\%}$ & $1.3223(8)_{0\%}$ &  
                    $0.03333(2)_{50\%}$  \\
\end{tabular}
\end{table}
Without choosing any particular value for them,
but simply averaging over all values between 0 and 10,
we obtain the results of Table \ref{stimeexp_1}. 
Note that we have not quoted any estimate of $\eta$ for $p=4$: in all cases,
some Pad\'e approximant was defective.
The quoted uncertainty, that expresses the variation of the estimates 
when changing $b_u$, $b_v$, and $r_u$, is very small,
and it is clear that it cannot be interpreted as a correct 
estimate of the error, since the variation with the order $p$
of the series is much larger. In Table \ref{stimeexp_1} we also report 
the estimates of the exponents corresponding to several different
values of $(\ub^*,\vb^*)$: beside the estimate (\ref{stimeustarvstar_2}),
we consider two values appearing in the first analysis of the 
fixed point position, those with the largest
and smallest value of $\ub^*$, when $b_u$ and $b_v$ vary in $[0,5]$. 
The dependence on $(\ub^*,\vb^*)$ is quite small, of the same 
order of the dependence on the order $p$.
As final estimate we quote the value obtained for $p=6$, 
using the estimate (\ref{stimeustarvstar_2}) for the fixed point. 
The error is estimated
by the difference between the results with $p=5$ and $p=6$. Therefore we have
\begin{equation}
\gamma = 1.313(14), \qquad\qquad 
\nu    = 0.668(6),  \qquad\qquad
\eta   = 0.0327(19).
\label{esponenti-stima1}
\end{equation}
Note that, within one error bar, all estimates of $\gamma$ and $\nu$
reported in Table \ref{stimeexp_1} are compatible with the results given above. 
Instead, the estimates of $\eta$ show a stronger dependence on the critical 
point, and {\em a priori}, since we do not know how reliable are the 
uncertainties reported in Eq. (\ref{stimeustarvstar_2}),
it is possible that the correct estimate is outside the confidence
interval reported above.

\begin{table}[tbp]
\caption{
Estimates of the critical exponents for the
RIM universality class obtained using the conformal Pad\'e-Borel 
method (first analysis).
All estimates correspond to the following choices of 
the parameters: $0\le b_u \le 10$, $2\le b_v \le 8$,
$-1.5\le \alpha_v \le 0.5$, $r_u=1,2$. The subscript indicates 
the percentage of defective Pad\'e approximants. The upper part of the 
table reports the estimates for $(\ub^*,\vb^*) = (-0.631,2.195)$ 
and different values of $p$ and $q$. The lower part reports 
estimates obtained setting $p=6$ and $q=4$ for several values of 
$(\ub^*,\vb^*)$.
}
\label{exponentsconformal1}
\begin{tabular}{lcccc}
  &  $p=5$, $q=3$   &  $p=5$, $q=4$ &  $p=6$, $q=3$ &  $p=6$, $q=4$ \\
\hline
$\gamma$ &  $1.336(18)_{40\%}$  & $1.308(19)_{40\%}$  &
            $1.305(15)_{50\%}$  & $1.338(21)_{0\%}$  \\
$\nu$    &  $0.673(11)_{44\%}$  & $0.660(10)_{44\%}$  &
            $0.657(7)_{54\%}$   & $0.676(11)_{0\%}$  \\
$\eta$   &  $0.0285(15)_{4\%}$  & $0.0288(17)_{46\%}$ &
            $0.0278(4)_{27\%}$  & $0.0279(5)_{31\%}$  \\
\hline\hline
   & ($-$0.647,2.215)   & ($-$0.615,2.175) &
     ($-$0.700,2.290)   & ($-$0.570,2.210) \\
\hline
$\gamma$ &  $1.342(21)_{0\%}$  & $1.334(21)_{0\%}$  &
            $1.371(23)_{0\%}$  & $1.349(21)_{0\%}$  \\
$\nu$    &  $0.678(11)_{0\%}$  & $0.674(11)_{0\%}$ & 
            $0.694(13)_{0\%}$  & $0.685(13)_{0\%}$  \\
$\eta$   &  $0.0277(5)_{31\%}$ & $0.0280(4)_{33\%}$ &
            $0.0270(12)_{28\%}$ & $0.0334(4)_{30\%}$ \\
\end{tabular}
\end{table}
   
We will now use the conformal Pad\'e-Borel method.
A first estimate is obtained considering approximants of the form
\begin{equation}
\widehat{E}_3(\cdot)(q,p;b_u,r_u;b_v,\alpha_v) = 
   E_2(\cdot)(q,p;b_u,r_u;\{b_n=b_v\},\{\alpha_n=\alpha_v\}),
\end{equation}
setting all $b_n$ equal to $b_v$ and $\alpha_n$ equal to $\alpha_v$.
The results show a tiny dependence on $b_u$, while no systematic difference is
observed between the approximants with $r_u=1$ and $r_u=2$. 
Therefore, we averaged over all non-defective results with $0\le b_u\le 10$ and 
$r_u = 1,2$. Then, we looked for optimal intervals 
$[\alpha_{\rm opt} - \Delta\alpha,\alpha_{\rm opt} + \Delta\alpha]$, 
$[b_{\rm opt} - \Delta b,b_{\rm opt} + \Delta b]$ for the parameters
$\alpha_v$ and $b_v$. They were determined by minimizing the 
discrepancies among the estimates corresponding to 
$(p,q) = (5,3)$, $(5,4)$, $(6,3)$, and $(6,4)$. Using $\Delta\alpha=1$ and 
$\Delta b = 3$ as we did before, we obtain 
$b_{\rm opt} = 5$ and $\alpha_{\rm opt} = -0.5$. The results corresponding
to this choice of parameters are reported in Table 
\ref{exponentsconformal1}. 
As final estimate we quote the value obtained for $p=6$ and $q=4$,
using the estimate (\ref{stimeustarvstar_2}) for the fixed point:
\begin{equation}
\gamma = 1.338(21), \qquad\qquad
\nu = 0.676(11),    \qquad\qquad
\eta = 0.0279(5).
\label{esponenti-stima2}
\end{equation}
For $\gamma$ and $\nu$ the estimates given above are compatible with all 
results of Table \ref{exponentsconformal1}. In particular, they are 
correct even if the error in Eq. (\ref{stimeustarvstar_2}) is 
underestimated. They are also in good agreement with the estimates 
obtained with the double Pad\'e-Borel transformation, 
cf. Eq. (\ref{esponenti-stima1}). On the other hand, it is not clear 
if the error on $\eta$ is reliable. Indeed, comparison with 
Eq. (\ref{esponenti-stima1}) may indicate that the correct value of $\eta$ 
is larger than what predicted by this analysis.

\begin{table}[tbp]
\caption{
Estimates of the critical exponents for the
RIM universality class obtained using the conformal Pad\'e-Borel method
(second analysis).
All estimates correspond to the following choices of 
the parameters: $0\le b_u \le 10$, $-3\le \delta_b \le 3$,
$-1\le \delta_\alpha \le 1$, $r_u=1,2$. The subscript indicates 
the percentage of defective Pad\'e approximants. The upper part of the 
table reports the estimates for $(\ub^*,\vb^*) = (-0.631,2.195)$ 
and different values of $p$ and $q$. The lower part reports 
estimates obtained setting $p=6$ and $q=4$ for several values of 
$(\ub^*,\vb^*)$.
}
\label{exponentsconformal2}
\begin{tabular}{lcccc}
  &  $p=5$, $q=3$   &  $p=5$, $q=4$ &  $p=6$, $q=3$ &  $p=6$, $q=4$ \\
\hline
$\gamma$ &  $1.335(16)_{42\%}$  & $1.329(15)_{67\%}$  &
            $1.322(7)_{47\%}$  & $1.321(8)_{56\%}$  \\
$\nu$    &  $0.684(3)_{45\%}$  & $0.682(3)_{72\%}$  &
            $0.682(1)_{33\%}$   & $0.681(1)_{67\%}$  \\
$\eta$   &  $0.0299(4)_{0\%}$  & $0.0299(15)_{0\%}$ &
            $0.0312(6)_{0\%}$  & $0.0313(5)_{0\%}$  \\
\hline\hline
   & ($-$0.647,2.215)   & ($-$0.615,2.175) &
     ($-$0.700,2.290)   & ($-$0.570,2.210) \\
\hline
$\gamma$ &  $1.323(8)_{67\%}$  & $1.318(8)_{67\%}$  &
            $1.332(10)_{67\%}$  & $1.331(7)_{67\%}$  \\
$\nu$    &  $0.683(1)_{68\%}$  & $0.680(1)_{65\%}$ & 
            $0.690(1)_{71\%}$  & $0.687(1)_{68\%}$  \\
$\eta$   &  $0.0314(5)_{0\%}$ & $0.0313(5)_{0\%}$ &
            $0.0315(5)_{0\%}$ & $0.0368(6)_{0\%}$ \\
\end{tabular}
\end{table}

As we did for the fixed point, we can also use the approximants 
$\widehat{E}_2$ defined in Eq. (\ref{def-approximantsE2}), optimizing
separately each coefficient. The results are reported in 
Table \ref{exponentsconformal2} and correspond to 
$0\le b_u\le 10$, $-3\le \delta_b\le 3$, $-1\le \delta_\alpha\le 1$
and $r_u=1,2$. As it can be seen from the very small
``errors" on the results, the dependence on $b_u$ is tiny and we have not 
tried to optimize this parameter. 
The results are reasonably stable with respect to
changes of $p$ and $q$ and also the dependence on the value of the 
fixed point is small. As final results from this analysis we quote
the values obtained with $p=6$ and $q=4$:
\begin{equation}
\gamma = 1.321(8), \qquad\qquad
\nu = 0.681(1),    \qquad\qquad
\eta = 0.0313(5).
\label{esponenti-stima3}
\end{equation}
We can compare these results with the previous estimates 
(\ref{esponenti-stima1}) and (\ref{esponenti-stima2}). The agreement is  
reasonable, although the quoted error on $\nu$ and $\eta$ is probably
underestimated. This is confirmed by the fact that the scaling 
relation $\gamma = (2 - \eta)\nu$ is not satisfied within error bars:
indeed, using the estimates of $\nu$ and $\eta$, we get $\gamma = 1.341(2)$.

We want now to obtain final estimates from the analyses given above. 
Since in the conformal Pad\'e-Borel method we make use of some 
additional information, the position of the singularity of the 
Borel transform, we believe this analysis to be the more reliable one.
As our finale estimate we have therefore considered the average 
between (\ref{esponenti-stima2}) and (\ref{esponenti-stima3}), fixing 
the error in such a way to include also the estimates 
(\ref{esponenti-stima1}). In this way we obtain
\begin{equation}
\gamma = 1.330(17), \qquad\qquad
\nu = 0.678(10),    \qquad\qquad
\eta = 0.030(3).
\label{esponenti-stimafinale}
\end{equation}
A check of these results is provided by the scaling relation 
$\gamma = \nu(2-\eta)$. Using the values of $\nu$ and $\eta$ 
we obtain $\gamma = 1.336 (20)$ in good agreement with the direct estimate.
A second check of these results is given by the 
inequalities $\nu \ge 2/3\approx 0.66667$ and 
$\gamma + 2\eta/3 \ge 4/3 \approx 1.33333$, that are clearly satisfied by
our estimates, e.g. $\gamma + 2 \eta/3 = 1.350(17)$.
Finally, we want to stress that our final estimates 
(\ref{esponenti-stimafinale}) are compatible with 
all results appearing in Tables
\ref{stimeexp_1}, \ref{exponentsconformal1}, and 
\ref{exponentsconformal2}, even those computed for $(\ub^*,\vb^*)$
largely different from the estimate (\ref{stimeustarvstar_2}).
Thus, we believe that our error estimates take properly into account the 
uncertainty on the position of the fixed point.

\begin{table}[tbp]
\caption{Field-theoretic 
estimates of the critical exponents for the 
RIM universality class. Here ``$d=3$ exp." denotes the massive scheme 
in three dimensions, ``MS exp." the minimal subtraction scheme 
without $\epsilon$-expansion. All perturbative results have been obtained 
by means of Pad\'e-Borel or Chisholm-Borel resummations, except the 
results of Ref. \protect\cite{Mayer-89} indicated by ``$\epsilon$W" 
obtained using the $\epsilon$-algorithm of Wynn and of this work.
}
\label{table_exponents}
\begin{tabular}{cccccc}
  &
\multicolumn{1}{c}{Method}&
\multicolumn{1}{c}{$\gamma$}&
\multicolumn{1}{c}{$\nu$}&
\multicolumn{1}{c}{$\eta$}&
\multicolumn{1}{c}{$\omega$} \\
\hline
This work               & $d=3$ exp. $O(g^6)$ &
   1.330(17) & 0.678(10) & 0.030(3) & 0.25(10) \\
\hline
Ref. \protect\cite{PS-99}, 2000        & $d=3$ exp. $O(g^5)$ &
        1.325(3) & 0.671(5)  & 0.025(10) & 0.32(6) \\
Ref. \protect\cite{Varnashev-99}, 2000 & $d=3$ exp. $O(g^4)$ & 
        1.336(2) & 0.681(12) & 0.040(11) & 0.31 \\
  & &   1.323(5) & 0.672(4)  & 0.034(10) & 0.33 \\
Ref. \protect\cite{FHY-99}, 1999       & $d=3$ exp. $O(g^4)$ &
         & & & 0.372(5) \\
Ref. \protect\cite{Mayer-89}, 1989  & $d=3$ exp. $O(g^4)$ & 
   1.321   & 0.671  &  & \\
Ref. \protect\cite{Mayer-89}, 1989  & $d=3$ exp. $O(g^4)$ $\epsilon$W & 
   1.318   & 0.668  &  & \\
Ref. \protect\cite{MSS-89}, 1989   & $d=3$ exp. $O(g^4)$ &
        1.326 & 0.670 & 0.034 & \\
\hline
Ref. \protect\cite{FHY-99,FHY-00}, 1999   & $d=3$ MS $O(g^4)$ &
    1.318   & 0.675 & 0.049  &  0.39(4)      \\
\hline
Ref. \protect\cite{NR-82}, 1982    & scaling field   &
    1.38   & 0.70  & & 0.42 \\
\end{tabular}
\end{table}

Using the scaling relations $\alpha = 2 - 3\nu$ and $\beta = 
{1\over2} \nu (1 + \eta)$ we have 
\begin{equation}
\alpha = -0.034(30), \qquad\qquad \beta = 0.349(5).
\end{equation}

For comparison we have also performed the direct analysis of the 
perturbative series, resumming the expansions for fixed 
$\vb^*/\ub^*$. In zero dimensions, these series are not Borel summable, 
and this is expected to be true in any dimension. However, for the 
short series we are considering, we can still hope to obtain 
reasonable results. We have used the same procedures described in 
Ref. \cite{CPV-00}, performing a conformal transformation and 
using $\ub_b(z)$ given in Eq. (\ref{bsingN0}) as 
position of the singularity. We obtain
\begin{eqnarray}
\ub^* = - 0.763(25), \qquad && \qquad \vb^* = 2.306(43);
\\
\gamma = 1.327(12), \qquad\qquad
\nu = 0.673(8),    \qquad && \qquad
\eta = 0.029(3),   \qquad\qquad
\omega = 0.34(11).
\end{eqnarray}
The estimate of the fixed point is very different from that 
computed before. This may indicate that the non-Borel summability causes
a large systematic error in this type of analysis. Probably the 
optimal truncation for the $\beta$-functions corresponds to shorter 
series. On the other hand, the critical exponents show a tiny dependence 
on the position of the fixed point. The estimates we obtain 
are in good agreement with our previous ones, indicating that 
the exponent series are much better behaved.

Let us now compare our results with previous field-theoretic determinations,
see Table \ref{table_exponents}. We observe a very good agrement with 
all the reported results. Note that our error bars on $\gamma$ and $\nu$
are larger than those previously quoted. We believe our uncertainties to 
be more realistic. Indeed, we have often found in this work, that 
Pad\'e-Borel estimates are insensitive to the parameters used in the analysis,
in particular to the parameter $b$ characterizing the Borel-Leroy
transform. Therefore, error estimates based on this criterion may underestimate
the uncertainty of the results. The perturbative results reported in Table 
\ref{table_exponents} correspond to the massive scheme in fixed dimension
$d=3$ and to the minimal subtraction scheme without $\epsilon$-expansion.
It should be noted that the latter scheme does not provide any 
estimate at five loops \cite{FHY-00}. Indeed, at this order the resummed 
$\beta$-function do not have any zero in the region $u<0$. This fact 
is probably related to the fact that the series which is analyzed is not 
Borel-summable. Therefore, perturbative expansions should have an 
optimal truncation beyond which the quality of the results worsens. 
For the $\beta$-functions in the minimal subtraction scheme, the optimal 
number of loops appears to be four. 
Two other methods have been used to compute the critical exponents:
the scaling-field method \cite{NR-82}
and the $\sqrt{\epsilon}$-expansion \cite{Khmelnitskii-75}. 
The former gives reasonable results, while the latter is unable to 
provide quantitative estimates of critical quantities, see e.g. 
Ref. \cite{FHY-00}. 
We can also compare our results with the recent Monte Carlo estimates of 
Ref. \cite{BFMMPR-98}. The agreement is quite good for $\gamma$ and $\nu$,
while our estimate of $\eta$ is slightly smaller, although still 
compatible within one error bar. This is not unexpected and appears
as a general feature of the $d=3$-expansion: indeed, also for the pure 
model, the estimate of $\eta$ obtained in the fixed-dimension expansion
is lower than the Monte Carlo and high-temperature results
(see Ref. \cite{CPRV-99} and references therein).

\subsection{The random $M$-vector model for $M\ge 2$} 

In this Section we consider the random vector model for $M\ge 2$. First, we
have studied the region $u < 0$, looking for a possible fixed point. 
We have not found any stable solution, in agreement with the general
arguments given in the introduction:
the mixed point is indeed located in the region $u>0$, and 
it is therefore irrelevant for the critical behavior of the dilute 
model. What remains to be checked is the stability of the 
$O(M)$ fixed point. As we mentioned in the introduction, a general argument 
predicts that this fixed point is stable; the random (cubic) perturbation 
introduces only subleading corrections with exponent 
$\omega = -\alpha_M/\nu_M$. This exponent can be easily computed from 
\begin{equation}
\omega = {\partial \beta_{\overline{u}} \over \partial\overline{u}}
  (0,\overline{v}),
\end{equation}
which is $N$-independent as expected. 

The analysis is identical to that performed for the stability of the 
Ising point in Ref.~\cite{CPV-00}. We use a conformal transformation 
and the large-order behavior of the series; then we fix the optimal
values $b_{\rm opt}$ and $\alpha_{\rm opt}$ of the parameters by 
requiring the estimates of $\overline{v}^*$ and $\omega$ to be 
stable with respect to the order of the series used. The errors were 
obtained varying $\alpha$ and $b$ in the intervals
$[\alpha_{\rm opt}-2,\alpha_{\rm opt}+2]$ and 
$[b_{\rm opt} - 3,b_{\rm opt} + 3]$. As in Ref.~\cite{CPV-00}, the final 
result is reported with an uncertainty corresponding to two 
standard deviations.

\begin{table}[tbp]
\caption{
Estimates of the subleading exponent $\omega$ at the $O(M)$-symmetric
fixed point. The last column reports the theoretical prediction
$-\alpha_M/\nu_M$.
}
\label{esponenteomega_Mvettore}
\begin{tabular}{lrrrrl}
\multicolumn{1}{c}{$M$}   &   
\multicolumn{1}{c}{$p=4$}  & 
\multicolumn{1}{c}{$p=5$}  & 
\multicolumn{1}{c}{$p=6$}  & 
\multicolumn{1}{c}{final}  &  
\multicolumn{1}{c}{$-\alpha_M/\nu_M$} \\
\hline
2     &  0.009(36)  &  0.003(9)  & 0.007(4)    & 0.007(8) &
              0.0192(6)  Ref. \cite{LSNCI-96} \\
 & & & & &    0.0223(24) Ref. \cite{CPRV-00}    \\
 & & & & &    0.0163(67) Ref. \cite{GZ-98} \\ 
\hline
3     &     0.142(19)  &  0.151(8)  & 0.156(5)    & 0.156(10) &
     0.172(14) Ref. \cite{GZ-98}   \\
 & & & & & 0.203(12) Ref. \cite{Butera-Comi_97} \\
\hline
4     &     0.274(35)  &  0.269(10) & 0.280(6)    & 0.280(12) &
     0.301(22) Ref. \cite{GZ-98}   \\
 & & & & & 0.360(16) Ref. \cite{Butera-Comi_97} \\
\hline
8     &     0.580(40)  &  0.563(13) & 0.586(8)    & 0.586(16) &
     0.683(11) Ref. \cite{Butera-Comi_97} \\
\end{tabular}
\end{table}

The final results for some values of $M$ are reported in Table 
\ref{esponenteomega_Mvettore}, together with estimates of 
the theoretical prediction $-\alpha_M/\nu_M$. For $M\ge 3$ these 
results clearly indicate that the $O(M)$-symmetric point is stable.
The results are somewhat lower than the theoretical prediction, especially
if we consider the high-temperature estimates of the critical exponents of 
Ref. \cite{Butera-Comi_97}. This is not surprising: indeed the 
estimates of the subleading exponents $\omega$ show in many cases 
discrepancies with estimates obtained by using other methods. 
This is probably connected to the non-analyticity of the 
$\beta$-function at the fixed point 
\cite{Parisi-80,Nickel-82,Nickel-91,MN-91,PV-gr-98}. A similar discrepancy,
although still well within a combined error bar, is observed for $M=2$.
In this case, we obtain $\omega > 0$, indicating that the fixed point 
is stable. The error however does not allow to exclude the opposite case.

\acknowledgements
We thank Victor Mart\'\i n-Mayor and
Giorgio Parisi for useful discussions.



\begin{references}
\bibitem{footnoteHedilute}
Ref. \cite{YC-97} reports $\nu=0.67(1)$ and Ref. \cite{ZR-99} 
finds that the exponent $\nu$ is compatible with 2/3. These
estimates agree with the best experimental result for the 
pure system $\nu = 0.67095(13)$ (obtained from the estimate 
of $\alpha$ using the hyperscaling relation $2 - \alpha = d \nu$)
\cite{LSNCI-96} and with the theoretical prediction
$\nu = 0.67166(55)$ \cite{CPRV-00}. There esist also experiments 
for ${}^4$He in aerogels, which, however, find larger values for
the exponent $\nu$ \cite{CBMWR-88,MMGA-91,CMR-96}. 
The current explanation of these results is that, in aerogels, 
the silica network is correlated to long distances, and therefore,
the Harris criterion and the model studied in this paper do not apply.
A simple model describing these materials has been studied in 
Ref. \cite{LT-89}.
\end{references}
\end{document}